\documentclass[twocolumn,showpacs,showkeys]{revtex4}
\usepackage{graphicx}
\usepackage{slashbox}
\input{psfig.sty}

\newcommand{\bastar}{\begin{eqnarray*}}
\newcommand{\eastar}{\end{eqnarray*}}
\newskip\humongous \humongous=0pt plus 1000pt minus 1000pt

\newif\ifdtup

\relax
\newcommand{\be}{\begin{equation}}
\newcommand{\ee}{\end{equation}}
\newcommand{\bea}{\begin{eqnarray}}
\newcommand{\eea}{\end{eqnarray}}

\newcommand{\pro}{\partial}
\newcommand{\n}{\hat n}

\newcommand{\dfrac}{\displaystyle\frac}
\newcommand{\ba}{\begin{array}}
\newcommand{\ea}{\end{array}}

\newcommand{\nn}{\nonumber}
\newcommand{\hn}{\hat n}
\begin{document}
\title{Topological Objects in Two-component Bose-Einstein Condensates}
\author{Y. M. Cho}
\email{ymcho@yongmin.snu.ac.kr}
\affiliation{Center for Theoretical Physics,
College of Natural Sciences, Seoul National University,
Seoul 151-742, Korea  \\
and \\
C.N.Yang Institute for Theoretical Physics, State University of
New York, Stony Brook, NY 11790, USA}
\author{Hyojoong Khim and Pengming Zhang}
\email{kimhj76@phya.snu.ac.kr}
\email{zhpm@phya.snu.ac.kr}
\affiliation{School of Physics,
College of Natural Sciences, \\
Seoul National University, Seoul 151-742, Korea}
\begin{abstract}
~~~~~We study the topological objects in two-component
Bose-Einstein condensates. We compare two competing theories
of two-component Bose-Einstein condensate, the popular
Gross-Pitaevskii theory and the recently proposed gauge theory of
two-component Bose-Einstein condensate which has an induced
vorticity interaction. We show that two theories produce very
similar topological objects, in spite of the obvious differences
in dynamics. Furthermore we show that the gauge theory of
two-component Bose-Einstein condensate, with the $U(1)$ gauge
symmetry, is remarkably similar to the Skyrme theory. Just like
the Skyrme theory the theory admits the non-Abelian
vortex, the helical vortex, and the vorticity knot. We
construct the lightest knot solution in two-component
Bose-Einstein condensate numerically,
and discuss how the knot can be constructed in the
spin-1/2 condensate of $^{87}{\rm Rb}$ atoms.
\end{abstract}
%\pacs{03.75.Fi, 05.30.Jp, 67.40.Vs, 74.72,-h}



\date{\today}
\maketitle

\section{Introduction}

Topological objects, in particular finite energy
topological objects, have played important roles in physics
\cite{abri,skyr}. In Bose-Einstein condensates (BEC) the best
known topological objects are the vortices, which have been widely
studied in the literature. Theoretically these vortices have
successfully been described by the Gross-Pitaevskii Lagrangian. On
the other hand, the recent advent of multi-component BEC, in
particular the spin-1/2 condensate of $^{87}{\rm Rb}$ atoms, has
widely opened an interesting possibility for us to construct
totally new topological objects in condensed matter physics. This
is because the multi-component BEC obviously has more interesting
non-Abelian structure which does not exist in ordinary
(one-component) BEC, and thus could admit new topological objects
which are absent in ordinary BEC \cite{exp1,exp2}. As importantly,
the multi-component BEC provides a rare opportunity to study the
dynamics of the topological objects theoretically. The dynamics of
multi-component BEC could be significantly different from that of
ordinary BEC. This is because the velocity field of the
multi-component BEC, unlike the ordinary BEC, in general has a
non-vanishing vorticity which could play an important role in the
dynamics of the multi-component BEC \cite{bec1}. So the
multi-component BEC provides an excellent opportunity for us to
study non-Abelian dynamics of the condensate theoretically and
experimentally.

The purpose of this paper is to discuss the non-Abelian dynamics of
two-component BEC. We first study the popular Gross-Pitaevskii
theory of two-component BEC, and compare the theory with the
recent gauge theory of two-component BEC which has a vorticity
interaction \cite{bec1}. We show that,
in spite of the obvious dynamical differences,
two theories are not much different physically. In particular,
they admit remarkably similar topological objects, the
helical vortex whose topology is fixed by $\pi_2(S^2)$ and the
vorticity knot whose topology is fixed by $\pi_3(S^2)$. Moreover
we show that the vorticity knot is nothing but the vortex ring
made of the helical vortex. Finally we show that the gauge theory
of two-component BEC is very similar to the theory of two-gap
superconductors, which implies that our analysis here can have an
important implication in two-gap superconductors.

A prototype non-Abelian knot is the Faddeev-Niemi knot in Skyrme
theory \cite{cho01,fadd1}. The vorticity knot in two-component BEC
turns out to be surprisingly similar to the Faddeev-Niemi knot. So it is
important for us to understand the Faddeev-Niemi knot first. The
Faddeev-Niemi knot is described by a non-linear sigma field $\hn$
(with $\hn^2=1$) which defines the Hopf mapping $\pi_3(S^2)$, the
mapping from the compactified space $S^3$ to the target space
$S^2$ of $\hn$, in which the preimage of any point in the target
space becomes a closed ring in $S^3$. When $\pi_3(S^2)$ becomes
non-trivial, the preimages of any two points in the target space
are linked, with the linking number fixed by the third homotopy of
the Hopf mapping. In this case the mapping is said to describe a
knot, with the knot quantum number identified by the linking
number of two rings. And it is this Hopf mapping that describes
the topology of the Faddeev-Niemi knot \cite{cho01,fadd1,sky3}.

In this paper we show that the vorticity knot in two-component BEC
has exactly the same topology as the Faddeev-Niemi knot. The only
difference is that here the vorticity knot in two-component BEC
has the extra dressing of the scalar field which represents the
density of the condensation.

The paper is organized as follows. In Section II we review the
Skyrme theory to emphasize its relevance in condensed matter
physics. In Section III we review the topological objects in
Skyrme theory in order to compare them with those in two-component
BEC. In Section IV we review the popular Gross-Pitaevskii theory
of two-component BEC, and show that the theory 
is closely related to Skyrme theory. 
In Section V we discuss the helical vortex
in Gross-Pitaevskii theory of two-component BEC, and show that it
is a twisted vorticity flux. In Section VI we discuss the gauge
theory of two-component BEC which includes the vorticity
interaction, and compare it with the Gross-Pitaevskii theory of
two-component BEC. In Section VII we discuss the helical vortex in
gauge theory of two-component BEC, and compare it with those in
Gross-Pitaevskii theory and Skyrme theory. We demonstrate that the
helical vortex in all three theories are remarkably similar to one
another. In Section VIII we present a numerical knot solution in
the gauge theory of two-component BEC, and show that it is 
nothing but the vortex ring
made of helical vorticity flux. Finally in Section IX we discuss
the physical implications of our result. In particular we
emphasize the similarity between the gauge theory of two-component
BEC and the theory of two-gap superconductor.

\section{Skyrme theory: A Review}

The Skyrme theory has long been interpreted as an effective field
theory of strong interaction with a remarkable success
\cite{prep}. However, it can also be interpreted as a theory of
monopoles, in which the monopole-antimonopole pairs are confined
through a built-in Meissner effect \cite{cho01,sky3}. This suggests that
the Skyrme theory could be viewed to describe a very interesting
condensed matter physics. Indeed the theory and the
theory of two-component BEC have many common features.
In particular, the topological objects that we
discuss here are very similar to those in Skyrme theory.
To understand this we review the Skyrme theory first.

Let $\omega$ and $\hat n$ (with ${\hat n}^2 = 1$) be the
massless scalar field and non-linear sigma field
in Skyrme theory, and let
\bea
&U = \exp (\dfrac{\omega}{2i} \vec \sigma \cdot \hat n)
= \cos \dfrac{\omega}{2} - i (\vec \sigma \cdot \hat n)
\sin \dfrac{\omega}{2}, \nn\\
&L_\mu = U\partial_\mu U^{\dagger}.
\label{su2}
\eea
With this one
can write the Skyrme Lagrangian as \cite{skyr}
\bea 
&{\cal L} = \dfrac{\mu^2}{4} {\rm tr} ~L_\mu^2 +
\dfrac{\alpha}{32}{\rm tr}
\left( \left[ L_\mu, L_\nu \right] \right)^2 \nn\\
&= - \dfrac{\mu^2}{4} \Big[ \dfrac{1}{2} (\partial_\mu \omega)^2
+2 \sin^2 \dfrac{\omega}{2} (\partial_\mu \hat n)^2 \Big] \nn\\
&-\dfrac{\alpha}{16} \Big[ \sin^2 \dfrac{\omega}{2} (\partial_\mu
\omega \partial_\nu \hat n 
-\partial_\nu \omega \partial_\mu \hat n)^2 \nn\\
&+4 \sin^4 \dfrac{\omega}{2} (\partial_\mu \hat n \times
\partial_\nu \hat n)^2 \Big],
\label{slag}
\eea
where $\mu$ and $\alpha$ are the coupling constants.
The Lagrangian has a hidden
local $U(1)$ symmetry as well as a global $SU(2)$ symmetry. From
the Lagrangian one has the following equations of motion
\bea
&\partial^2 \omega -\sin\omega (\partial_\mu \hat n)^2
+\dfrac{\alpha}{8 \mu^2} \sin\omega (\partial_\mu \omega
\partial_\nu \hat n -\partial_\nu \omega \partial_\mu \hat n)^2 \nn\\
&+\dfrac{\alpha}{\mu^2} \sin^2 \dfrac{\omega}{2}
\partial_\mu \big[ (\partial_\mu \omega \partial_\nu \hat n
-\partial_\nu \omega \partial_\mu \hat n)
\cdot \partial_\nu \hat n \big] \nn\\
&- \dfrac{\alpha}{\mu^2} \sin^2 \dfrac{\omega}{2} \sin\omega
(\partial_\mu \hat n \times
\partial_\nu \hat n)^2 =0, \nn \\
&\partial_\mu \Big\{\sin^2 \dfrac{\omega}{2}  \hat n \times
\partial_\mu \hat n \nn\\
&+ \dfrac{\alpha}{4\mu^2} \sin^2 \dfrac{\omega}{2}
\big[ (\partial_\nu \omega)^2 \hat n \times \partial_\mu \hat n
-(\partial_\mu \omega \partial_\nu \omega) \hat n \times
\partial_\nu \hat n \big] \nn\\
&+\dfrac{\alpha}{\mu^2} \sin^4 \dfrac{\omega}{2} (\hat n \cdot
\partial_\mu \hat n \times
\partial_\nu \hat n) \partial_\nu \hat n \Big\}=0.
\label{skeq1}
\eea
Notice that the second equation can be interpreted as a
conservation of $SU(2)$ current, which of course is a simple
consequence of the global $SU(2)$ symmetry of the theory.

With the spherically symmetric ansatz
\bea
\omega = \omega (r),
~~~~~\hat n = \hat r,
\label{skans1}
\eea
(3) is reduced to
\bea
&\dfrac{d^2 \omega}{dr^2} +\dfrac{2}{r} \dfrac{d\omega}{dr}
-\dfrac{2\sin\omega}{r^2} +\dfrac{2\alpha}{\mu^2}
\Big[\dfrac{\sin^2 (\omega/2)}{r^2}
\dfrac{d^2 \omega}{dr^2} \nn\\
&+\dfrac{\sin\omega}{4 r^2} (\dfrac{d\omega}{dr})^2
-\dfrac{\sin\omega \sin^2 (\omega /2)}{r^4} \Big] =0.
\label{skeq2}
\eea
Imposing the boundary condition
\bea
\omega(0)=2\pi,~~~~~\omega(\infty)= 0,
\label{skbc}
\eea
one can
solve the Eq. (\ref{skeq2}) and obtain the well-known skyrmion
which has a finite energy. The energy of the skyrmion is given by
\bea
&E = \dfrac{\pi}{2} \mu^2 \int^{\infty}_{0} \bigg\{\left(r^2
+ \dfrac{2\alpha}{\mu^2}
\sin^2{\dfrac{\omega}{2}}\right)\left(\dfrac{d\omega}{dr}\right)^2 \nn\\
&+8 \left(1 + \dfrac{\alpha}{2\mu^2~r^2}\sin^2{\dfrac{\omega}{2}}
\right)
\sin^2 \dfrac{\omega}{2} \bigg\} dr \nn\\
&= \pi {\sqrt \alpha} \mu \dfrac{}{} \int^{\infty}_{0} \Big[x^2
\left(\dfrac{d\omega}{dx}\right)^2
+ 8 \sin^2{\dfrac{\omega}{2}} \Big] dx \nn\\
&\simeq 73~{\sqrt \alpha} \mu.
\label{sken}
\eea
where $x=\mu/\sqrt \alpha$ is a dimensionless variable. 
Furthermore, it carries the baryon number \cite{skyr,prep}
\bea
&Q_s = \dfrac{1}{24\pi^2} \int
\epsilon_{ijk} ~{\rm tr} ~(L_i L_j L_k) d^3r=1,
\label{bn}
\eea
which represents the non-trivial homotopy
$\pi_3(S^3)$ of the mapping from the compactified space $S^3$ to
the $SU(2)$ space $S^3$ defined by $U$ in (\ref{su2}).

A remarkable point of (\ref{skeq1}) is that
\bea
\omega=\pi,
\eea
becomes a classical solution, independent of $\hn$ \cite{cho01}.
So restricting $\omega$ to $\pi$, one can reduce the Skyrme
Lagrangian (\ref{slag}) to the Skyrme-Faddeev Lagrangian
\bea
{\cal L} \rightarrow -\dfrac{\mu^2}{2} (\partial_\mu \hat
n)^2-\dfrac{\alpha}{4}(\partial_\mu \hat n \times
\partial_\nu \hat n)^2,
\label{sflag}
\eea
whose equation of motion is given by
\bea
&\hn \times \partial^2 \hn + \dfrac{\alpha}{\mu^2} (\partial_\mu
H_{\mu\nu}) \partial_\nu \hn = 0, \nn\\
&H_{\mu\nu} = \hn \cdot (\partial_\mu \hn \times \partial_\nu \hn)
=\partial_\mu C_\nu - \partial_\nu C_\mu.
\label{sfeq}
\eea
Notice that $H_{\mu\nu}$ admits a potential $C_\mu$ because it forms a
closed two-form. Again the equation can be viewed as a
conservation of $SU(2)$ current,
\bea
\partial_\mu (\hn \times \partial_\mu \hn
+\dfrac{\alpha}{\mu^2} H_{\mu\nu}\partial_\nu \hn) = 0.
\eea
It is this equation that allows not only the baby skyrmion and the
Faddeev-Niemi knot but also the non-Abelian monopole
\cite{cho01,sky3}.

\section{Topological Objects in Skyrme theory}

The Lagrangian (\ref{sflag}) has non-Abelian monopole solutions
\cite{cho01}
\bea
\hn = \hat r,
\label{mono1}
\eea
where $\hat r$
is the unit radial vector. This becomes a solution of (\ref{sfeq})
except at the origin, because
\bea
&\partial^2 \hat r = - \dfrac
{2}{r^2} \hat r, ~~~~\partial_\mu H_{\mu\nu} =0.
\eea
This is very similar to the well-known Wu-Yang monopole
in $SU(2)$ QCD \cite{cho01,prd80}. It has the magnetic charge
\bea
&Q_m = \dfrac{1}{8\pi} \int
\epsilon_{ijk} H_{ij} d\sigma_k=1,
\label{smqn}
\eea
which
represents the non-trivial homotopy $\pi_2(S^2)$ of the mapping
from the unit sphere $S^2$ centered at the origin in space to the
target space $S^2$.

The above exercise tells that we
can identify $H_{\mu\nu}$ a magnetic field and $C_\mu$ the
corresponding magnetic potential. As importantly this tells that
the skyrmion is nothing but a monopole dressed by the scalar
field $\omega$, which makes the energy of the skyrmion
finite \cite{cho01}.

\begin{figure}[t]
\includegraphics[scale=0.5]{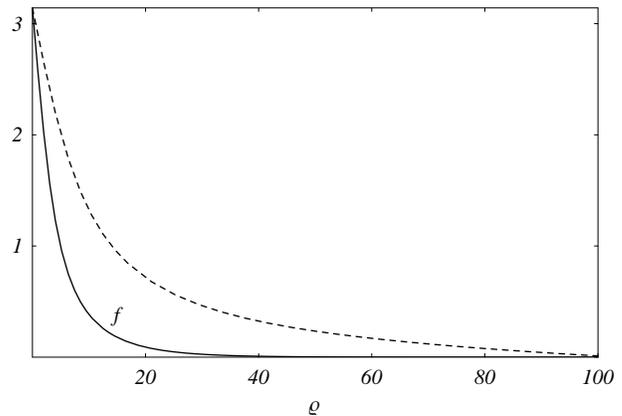}
\caption{The baby skyrmion (dashed line) with $m=0,n=1$ and the
helical baby skyrmion (solid line) with $m=n=1$ in Skyrme theory.
Here $\varrho$ is in the unit ${\sqrt \alpha}/\mu$
and $k=0.8~\mu/{\sqrt \alpha}$.}
\label{hbs}
\end{figure}

It has been well-known that the Skyrme theory has a vortex
solution known as the baby skyrmion \cite{piet}. Moreover, the
theory also has a twisted vortex solution, the helical baby
skyrmion \cite{sky3}. To construct the desired helical vortex
let $(\varrho,\varphi,z)$ be the cylindrical coordinates, and choose
the ansatz
\bea
&\n=\Bigg(\matrix{\sin{f(\varrho)}\cos{(n\varphi+mkz)} \cr
\sin{f(\varrho)}\sin{(n\varphi+mkz)} \cr \cos{f(\varrho)}}\Bigg).
\label{hvans}
\eea
With this we have (up to a gauge transformation)
\bea
&C_\mu
=-\big(\cos{f} +1\big) (n\pro_\mu \varphi+mk \pro_\mu z),
\eea
and can reduce the equation (\ref{sfeq}) to
\bea
&\Big(1+\dfrac{\alpha}{\mu^2}(\dfrac{n^2}{\varrho^2}+m^2 k^2)
\sin^2{f}\Big) \ddot{f} \nn\\&+ \Big( \dfrac{1}{\varrho}
+\dfrac{\alpha}{\mu^2}(\dfrac{n^2}{\varrho^2}+m^2 k^2) \dot{f}
\sin{f}\cos{f} \nn\\
&- \dfrac{\alpha}{\mu^2}\dfrac{1}{\varrho}
(\dfrac{n^2}{\varrho^2}-m^2 k^2) \sin^2{f} \Big) \dot{f} \nn\\
&- (\dfrac{n^2}{\varrho^2}+m^2 k^2) \sin{f}\cos{f}=0.
\label{hveq}
\eea
So with the boundary condition
\bea
f(0)=\pi,~~f(\infty)=0,
\label{bc}
\eea
we obtain the non-Abelian vortex solutions shown
in Fig.~\ref{hbs}. Notice that, when $m=0$, the solution describes
the well-known baby skyrmion. But when $m$ is not
zero, it describes a helical vortex which is periodic in
$z$-coordinate \cite{sky3}. In this case, the vortex has a non-vanishing
magnetic potential $C_\mu$ not only around the vortex but also
along the $z$-axis.

\begin{figure}[t]
\includegraphics[scale=0.5]{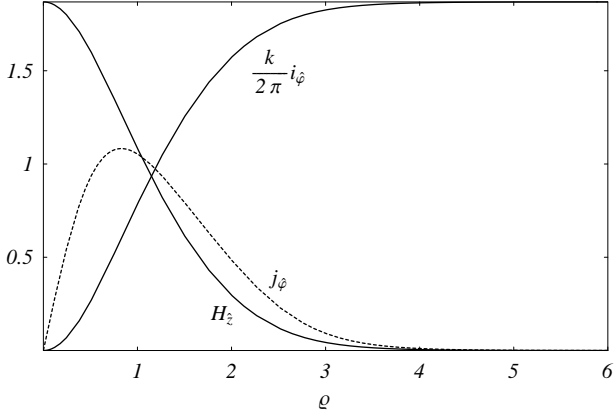}
\caption{The supercurrent $i_{\hat \varphi}$ (in one period
section in $z$-coordinate) and corresponding magnetic field
$H_{\hat z}$ circulating around the cylinder of radius $\varrho$
of the helical baby skyrmion with $m=n=1$. Here $\varrho$
is in the unit ${\sqrt \alpha}/\mu$ and $k=0.8~\mu/{\sqrt \alpha}$.
The current density $j_{\hat \varphi}$ is represented by the dotted line.}
\label{skyiphi}
\end{figure}

Obviously the helical vortex has the helical magnetic field
made of
\bea
&H_{\hat{z}}=\dfrac{1}{\varrho}H_{\varrho\varphi}
=\dfrac{n}{\varrho}\dot{f}\sin{f}, \nn\\
&H_{\hat{\varphi}}=-H_{\varrho z}= - mk \dot{f}\sin{f},
\eea
which
gives two quantized magnetic fluxes. It has a quantized magnetic
flux along the $z$-axis
\bea
&\phi_{\hat z} = \dfrac {}{}\int
H_{\varrho\varphi} d\varrho d\varphi = - 4\pi n,
\label{nqn1}
\eea
and a quantized magnetic flux around the $z$-axis (in one period
section from $0$ to $2\pi/k$ in $z$-coordinate)
\bea
&\phi_{\hat
\varphi} = -\dfrac {}{}\int H_{\varrho z} d\varrho dz = 4\pi m.
\label{mqn1}
\eea
Furthermore they are linked since $\phi_{\hat
z}$ is surrounded by $\phi_{\hat \varphi}$. This point will be
very important later when we discuss the knot.

\begin{figure}[t]
\includegraphics[scale=0.5]{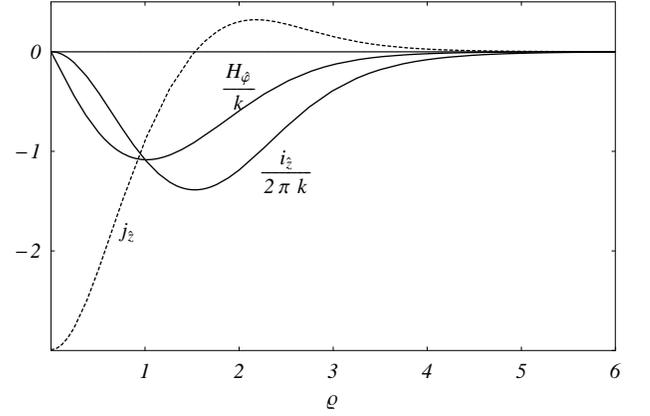}
\caption{The supercurrent $i_{\hat z}$ and corresponding magnetic
field $H_{\hat \varphi}$ flowing through the disk of radius
$\varrho$ of the helical baby skyrmion with $m=n=1$. Here $\varrho$
is in the unit ${\sqrt \alpha}/\mu$ and $k=0.8~\mu/{\sqrt \alpha}$.
The current density $j_{\hat z}$ is represented by the dotted line.}
\label{skyiz}
\end{figure}

The vortex solutions implies the existence of Meissner effect in
Skyrme theory which confines the magnetic flux of the vortex
\cite{sky3}. To see how the Meissner effect comes about, notice
that due to the $U(1)$ gauge symmetry the Skyrme theory has a
conserved current,
\bea
&j_\mu = \pro_\nu H_{\mu\nu},~~~~~\pro_\mu
j_\mu = 0.
\eea
So the magnetic flux of the vortex can be thought
to come from the helical electric current density
\bea
&j_\mu =- \sin f \Big[n \big(\ddot f + \dfrac{\cos f}{\sin f}
\dot f^2 - \dfrac{1}{\varrho} \dot f \big) \partial_{\mu}\varphi \nn\\
&- mk \big(\ddot f + \dfrac{\cos f}{\sin f} \dot f^2 +
\dfrac{1}{\varrho} \dot f \big) \partial_{\mu}z \Big].
\label{cc}
\eea
This produces the currents $i_{\hat\varphi}$ (per one period
section in $z$-coordinate from $z=0$ to $z=2\pi/k$) around the
$z$-axis
\bea
&i_{\hat\varphi} = - n \dfrac{}{}
\int_{\varrho=0}^{\varrho=\infty} \int_{z=0}^{z=2\pi/k}
\sin f \big(\ddot f + \dfrac{\cos f}{\sin f} \dot f^2 \nn\\
&- \dfrac{1}{\varrho} \dot f \big)  \dfrac{d\varrho}{\varrho} dz
=\dfrac{2 \pi n}{k}\dfrac{\sin{f}}{\varrho}\dot f
\Bigg|_{\varrho=0}^{\varrho=\infty} \nn\\
&=- \dfrac{2 \pi n}{k} \dot f^2(0),
\eea
and $i_{\hat z}$ along
the $z$-axis
\bea
&i_{\hat z} =- mk \dfrac{}{}
\int_{\varrho=0}^{\varrho=\infty} \sin f \big(\ddot f +
\dfrac{\cos f}{\sin f} \dot f^2
+ \dfrac{1}{\varrho} \dot f \big) \varrho d\varrho d\varphi \nn\\
&=- 2 \pi mk \varrho \dot f \sin{f}
\Bigg|_{\varrho=0}^{\varrho=\infty} =0.
\eea
Notice that, even
though $i_{\hat z}=0$, it has a non-trivial current density which
generates the net flux $\phi_{\hat \varphi}$.

The helical magnetic fields and currents are shown in
Fig.~\ref{skyiphi} and Fig.~\ref{skyiz}. Clearly the helical
magnetic fields are confined along the $z$-axis, confined by the
helical current. This is nothing but the Meissner effect, which
confirms that the Skyrme theory has a built-in mechanism for the
Meissner effect. 

The helical vortex will become unstable and decay to the untwisted
baby skyrmion unless the periodicity condition is enforced by
hand. In this sense it can be viewed as unphysical. But for our
purpose it plays a very important role, because it guarantees the
existence of the Faddeev-Niemi knot in Skyrme theory
\cite{cho01,sky3}. This is because we can naturally enforce the
periodicity condition of the helical vortex making it a vortex
ring by smoothly bending and connecting two periodic ends
together. In this case the periodicity condition is automatically
implemented, and the vortex ring becomes a stable knot.

The knot topology is described by the non-linear sigma field $\hat
n$, which defines the Hopf mapping from the compactified space
$S^3$ to the target space $S^2$. When the preimages of two points
of the target space are linked, the mapping $\pi_3(S^2)$
becomes non-trivial. In
this case the knot quantum number of $\pi_3(S^2)$ is given by the
Chern-Simon index of the magnetic potential $C_\mu$,
\bea
&Q_k = \dfrac{1}{32\pi^2} \int \epsilon_{ijk} C_i H_{jk} d^3x
= mn.
\label{kqn}
\eea
Notice that the knot quantum number can
also be understood as the linking number of two magnetic fluxes of
the vortex ring. This is because the vortex ring carries two
magnetic fluxes linked together, $m$ unit of flux passing through
the disk of the ring and $n$ unit of flux passing along the ring,
whose linking number becomes $mn$. This linking number is
described by the Chern-Simon index of the magnetic potential
\cite{sky3}.

The knot has both topological and dynamical stability. Obviously
the knot has a topological stability, because two flux rings
linked together can not be disconnected by any smooth deformation
of the field.

The dynamical stability follows from the fact that the 
supercurrent (\ref{cc}) has two components, the one moving along
the knot and the other moving around the knot tube. Clearly the
current moving along the knot generates an angular momentum around
the $z$-axis which provides the centrifugal force preventing the
vortex ring to collapse. Put it differently, the current generates
the $m$ unit of the magnetic flux trapped in the knot disk which
can not be squeezed out. And clearly, this flux provides a
stabilizing repulsive force which prevent the collapse of the
knot. This is how the knot acquires the dynamical stability. 
It is this remarkable interplay between topology and dynamics 
which assures the existence of the stable knot in 
Skyrme theory \cite{sky3}.

One could estimate the energy of the knot. Theoretically it has
been shown that the knot energy has the following bound
\cite{ussr}
\bea
c~\sqrt{\alpha}~\mu~Q^{3/4} \leq E_Q \leq
C~\sqrt{\alpha}~\mu~Q^{3/4}, \label{ke}
\eea
where $c=8\pi^2\times 3^{3/8}$ and $C$ is an unknown constant
not smaller than
$c$. This suggests that the knot energy is proportional to
$Q^{3/4}$. Indeed numerically, one finds \cite{batt2}
\bea
E_Q \simeq 252~\sqrt{\alpha}~\mu~Q^{3/4},
\label{nke}
\eea
up to $Q=8$. What is remarkable here is the sub-linear
$Q$-dependence of the energy. This means that a knot with large $Q$
can not decay to knots with smaller $Q$.

\section{Gross-Pitaevski Theory of Two-component BEC: A Review}

The creation of the multi-component Bose-Einstein condensates of
atomic gases has widely opened new opportunities for us to study
the topological objects experimentally which so far have been only
of theoretical interest. This is because the multi-component BEC
can naturally represent a non-Abelian structure, and thus can
allow far more interesting topological objects. Already the
vortices have successfully been created with different methods in
two-component BECs \cite{exp1,exp2}. But theoretically
the multi-component BEC has not been well-understood.
In particular, it need to be clarified how different
the vortices in multi-component BEC are from the
well-known vortices in the single-component BEC. This is an
important issue, because the new condensates could have a new
interaction, the vorticity interaction, which is absent in
single-component BECs. So in the following we first discuss 
the vortex in the popular Gross-Pitaevskii theory of 
two-component BEC, and compare it with that in gauge theory of 
two-component BEC which has been proposed
recently \cite{bec1}.

Let a complex doublet $\phi=(\phi_1,\phi_2)$ be the two-component
BEC, and consider the non-relativistic two-component
Gross-Pitaevskii Lagrangian \cite{ruo,batt1,gar,met}
\bea
&{\cal L} = i \dfrac {\hbar}{2} \Big[\big(\phi_1^\dag ( \partial_t \phi_1)
-( \partial_t \phi_1)^\dag \phi_1 \big) \nn\\
&+ \big(\phi_2^\dag ( \partial_t \phi_2) -( \partial_t
\phi_2)^\dag \phi_2 \big)  \Big]
- \dfrac {\hbar^2}{2M} (|\partial_i \phi_1|^2 + |\partial_i \phi_2|^2) \nn\\
& + \mu_1 \phi_1^\dag \phi_1 + \mu_2 \phi_2^\dag \phi_2
- \dfrac {\lambda_{11}}{2} (\phi_1^\dag \phi_1)^2 \nn\\
&- \lambda_{12} (\phi_1^\dag \phi_1)(\phi_2^\dag \phi_2) - \dfrac
{\lambda_{22}}{2} (\phi_2^\dag \phi_2)^2,
\label{gplag1}
\eea
where $\mu_i$ are the chemical potentials and $\lambda_{ij}$ 
are the quartic coupling constants which are determined 
by the scattering lengths $a_{ij}$
\bea
\lambda_{ij}=\dfrac{4\pi {\hbar}^2}{M} a_{ij}.
\eea
The Lagrangian (\ref{gplag1}) is a straightforward generalization of
the single-component Gross-Pitaevskii Lagrangian to the
two-component BEC. Notice that here we have neglected the trapping
potential. This is justified if the range of the trapping
potential is much larger than the size of topological objects we
are interested in, and this is what we are assuming here. Clearly
the Lagrangian has a global $U(1)\times U(1)$ symmetry.

One could simplify the Lagrangian (\ref{gplag1}) noticing the fact
that experimentally the scattering lengths often have the same
value. For example, for the spin $1/2$ condensate of $^{87}{\rm
Rb}$ atoms, all $a_{ij}$ have the same value of about $5.5~nm$
within $3~\%$ or so \cite{exp1,exp2}. In this case one may safely
assume
\bea
\lambda_{11} \simeq \lambda_{12} \simeq \lambda_{22}
\simeq \bar \lambda.
\label{qint}
\eea
With this assumption
(\ref{gplag1}) can be written as
\bea
&{\cal L} = i\dfrac
{\hbar}{2} \Big[\phi^\dag (\partial_t \phi) -(\partial_t
\phi)^\dag \phi \Big]
- \dfrac {\hbar^2}{2M} |\partial_i \phi|^2 \nn\\
&-\dfrac{\bar \lambda}{2} \big(\phi^\dag \phi -\dfrac{\mu}{\bar
\lambda} \big)^2 - \delta \mu \phi_2^\dag \phi_2,
\label{gplag2}
\eea
where
\bea
\mu=\mu_1,~~~~~\delta \mu = \mu_1-\mu_2.
\eea
Clearly the Lagrangian has a global $U(2)$ symmetry when $\delta
\mu=0$. So the $\delta \mu$ interaction can be understood to be
the symmetry breaking term which breaks the global $U(2)$ symmetry
to $U(1)\times U(1)$. Physically $\delta \mu$ represents the
difference of the chemical potentials between $\phi_1$ and
$\phi_2$ (Here one can always assume $\delta \mu \geq 0$ without
loss of generality), so that it vanishes when the two condensates
have the same chemical potential. Even when they differ the
difference could be small, in which case the symmetry breaking
interaction could be treated perturbatively. This tells that the
theory has an approximate global $U(2)$ symmetry, even in the
presence of the symmetry breaking term \cite{bec5}. This is why it
allows a non-Abelian topological objects.

Normalizing $\phi$ to $(\sqrt{2M}/\hbar)\phi$ and parametrizing
it by 
\bea
\phi = \dfrac {1}{\sqrt 2} \rho \zeta, 
~~~(|\phi|=\dfrac {1}{\sqrt 2} \rho,~\zeta^\dag
\zeta = 1)
\label{phi}
\eea
we obtain the following Hamiltonian from the Lagrangian (\ref{gplag2}) 
in the static limit (in the natural unit $c=\hbar=1$),
\bea
&{\cal H} =  \dfrac {1}{2} (\pro_i \rho)^2 +
\dfrac {1}{2} \rho^2 |\pro_i \zeta|^2
+ \dfrac{\lambda}{8} (\rho^2-\rho_0^2)^2 \nn\\
&+ \dfrac{\delta \mu^2}{2} \rho^2 \zeta_2^*\zeta_2,
\label{gpham1}
\eea
where
\bea
&\lambda=4M^2 \bar
\lambda,~~~~~\rho_0^2=\dfrac{4\mu M}{\lambda}, 
~~~~~\delta \mu^2=2M \delta \mu.
\eea
Minimizing the Hamiltonian we have
\bea
& \pro^2 \rho - |\pro_i \zeta|^2 \rho
=\Big (\dfrac{\lambda}{2} (\rho^2-\rho_0^2)
+ \delta \mu^2 (\zeta_2^* \zeta_2) \Big) \rho, \nn\\
&\Big\{(\pro^2 - \zeta^\dag \pro^2 \zeta) + 2 \dfrac {\pro_i
\rho}{\rho}(\pro_i - \zeta^\dag \pro_i\zeta) \nn\\
&+\delta \mu^2 (\zeta_2^* \zeta_2) \Big\} \zeta_1 = 0, \nn\\
&\Big\{(\pro^2 - \zeta^\dag \pro^2 \zeta) + 2 \dfrac {\pro_i
\rho}{\rho}(\pro_i - \zeta^\dag \pro_i\zeta) \nn\\
&-\delta \mu^2 (\zeta_1^* \zeta_1) \Big\} \zeta_2 = 0, \nn\\
&\zeta^\dag \pro_i(\rho^2\pro_i\zeta) 
-\pro_i(\rho^2\pro_i\zeta^\dag) \zeta =0.
\label{gpeq1}
\eea
The equation is closely related to 
the equation (\ref{sfeq}) we have in Skyrme theory, 
although on the surface it appears totally different from
(\ref{sfeq}). To show this we let
\bea
&\hn=\zeta^{\dagger} \vec \sigma \zeta, \nn\\
&C_\mu= -2i \zeta^{\dagger} \pro_\mu \zeta,
\label{hn}
\eea
and find 
\bea
&(\pro_\mu \hn)^2 = 4 \Big( |\pro_\mu \zeta|^2
- |\zeta^\dag \pro_\mu \zeta|^2\Big)
=4 |\pro_\mu \zeta|^2 - C_\mu^2, \nn\\
&\hn \cdot (\pro_\mu \hn \times \pro_\nu \hn) = -2i (\pro_\mu
\zeta^\dag \pro_\nu \zeta
- \pro_\nu \zeta^\dag \pro_\mu \zeta) \nn\\
&=\pro_\mu C_\nu - \pro_\nu C_\mu
= H_{\mu\nu}.
\label{nid}
\eea
Notice that here $H_{\mu\nu}$ is precisely the closed two-form
which appears in (\ref{sfeq}).
Moreover, from (\ref{hn}) we have the identity
\bea
&\Big[\pro_\mu +\dfrac{1}{2i}(C_\mu \hn 
-\hn \times \pro_\mu \hn) \cdot \vec \sigma \Big] \zeta =0.
\label{cid}
\eea
This identity plays an important role in non-Abelian 
gauge theory, which shows that there exists a unique $SU(2)$ 
gauge potential which parallelizes the doublet 
$\zeta$ \cite{prd80}. For our purpose this allows us 
to rewrite the equation of the doublet $\zeta$ 
in (\ref{gpeq1}) into a completely different form. 

Indeed with the above identities we can express (\ref{gpeq1}) 
in terms of $\hn$ and $C_\mu$. With (\ref{nid}) the first equation of 
(\ref{gpeq1}) can be written as
\bea
& \pro^2 \rho - \dfrac{1}{4}\big[(\pro_i \hn)^2 + C_i^2 \big] \rho
=\Big (\dfrac{\lambda}{2} (\rho^2-\rho_0^2) \nn\\
&+ \delta \mu^2 (\zeta_2^* \zeta_2) \Big) \rho. 
\eea
Moreover, with (\ref{cid}) the second and third equations of
(\ref{gpeq1}) can be expressed as
\bea
&\dfrac{1}{2i} \Big(A+\vec B \cdot \vec \sigma \Big) \zeta =0,  \nn\\
&A= \pro_i C_i+2\dfrac{\pro_i \rho}{\rho} C_i
+i (2 \zeta_2^* \zeta_2 - 1) \delta \mu^2, \nonumber \\                       
&\vec{B}= \hn \times \pro^2 \hn 
+2\dfrac{\pro_i \rho}{\rho} \hn \times \pro_i \hn
-C_i \pro_i \hn \nn\\
&- (\pro_i C_i +2\dfrac{\pro_i \rho}{\rho} C_i) \hn
+i \delta \mu ^2 \hat k,
\label{gpeq2b1}
\eea
where $\hat k=(0,0,1)$. This is equivalent to 
\bea
&A+ \vec B \cdot \hn=0, \nn\\
& \hn \times \vec B - i \hn \times (\hn \times \vec B) =0,
\eea
so that (\ref{gpeq2b1}) is written as
\bea
&\vec{n}\times \partial ^2\vec{n}+2\dfrac{\partial _i\rho }\rho                 
\vec{n}\times \partial_i\vec{n}-C_i\partial _i\vec{n} \nonumber \\              &=\delta \mu ^2 \hat k \times \hn.                     
\label{gpeq2b2}
\eea
Finally, the last equation of (\ref{gpeq1}) is written as
\bea
\partial_i (\rho^2 C_i) = 0,
\eea
which tells that $\rho^2 C_i$ is 
solenoidal (i.e., divergenceless). So we can always 
replace $C_i$ with another field $B_i$
\bea
&C_i= \dfrac {1}{\rho^2} \epsilon_{ijk} \pro_j B_k
=-\dfrac {1}{\rho^2} \pro_i G_{ij}, \nn\\
&G_{ij}=\epsilon_{ijk} B_k,
\eea  
and express (\ref{gpeq2b2}) as
\bea
&\hn \times \pro^2 \hn + 2\dfrac{\pro_i \rho}{\rho}
\hn \times \pro_i \hn + \dfrac{1}{\rho^2} 
\pro_i G_{ij} \pro_j \hn \nn\\
&=\delta \mu ^2\hat k \times \vec{n}.        
\eea
With this (\ref{gpeq1}) can now be written as
\bea
& \pro^2 \rho - \dfrac{1}{4} \big[(\pro_i \hn)^2 + C_i^2 \big] \rho
=\Big (\dfrac{\lambda}{2} (\rho^2-\rho_0^2) \nn\\
&+ \delta \mu^2 (\zeta_2^* \zeta_2) \Big) \rho, \nn\\
&\hn \times \pro^2 \hn + 2\dfrac{\pro_i \rho}{\rho}
\hn \times \pro_i \hn + \dfrac{1}{\rho^2} \pro_i G_{ij} \pro_j \hn 
=\delta \mu ^2\hat k \times \vec{n}, \nn\\        
&\pro_i G_{ij}= - \rho^2 C_j.
\label{gpeq2}
\eea
This tells that (\ref{gpeq1}) can be transformed to
a completely different form which has a clear physical meaning.
The last equation tells that the theory has a conserved $U(1)$
current $j_\mu$,
\bea
j_\mu=\rho^2 C_\mu, 
\eea
which is nothing but the Noether current
of the global $U(1)$ symmetry of the Lagrangian (\ref{gplag2}).
The second equation tells that the theory has another 
partially conserved $SU(2)$ Noether current $\vec j_\mu$,
\bea
\vec j_\mu= \rho^2 \hn \times \pro_\mu \hn 
-\rho^2 C_\mu \hn, 
\eea
which comes from the approximate $SU(2)$ symmetry 
of the theory broken by the $\delta \mu$ term. 
It also tells that the theory has one more $U(1)$ current
\bea
k_\mu= \hat k \cdot \vec j_\mu,
\eea
which is conserverd even when $\delta \mu$ is not zero.
This is because, when $\delta \mu$ is not zero, the $SU(2)$ symmetry
is broken down to a $U(1)$.

More importantly this shows 
that (\ref{gpeq1}) is not much different from the equation 
(\ref{sfeq}) in Skyrme theory. Indeed in the absence of $\rho$,
(\ref{sfeq}) and (\ref{gpeq2}) acquire an identical form 
when $\delta \mu^2=0$, except that here $H_{ij}$ is 
replaced by $G_{ij}$.
This reveals that the Gross-Pitaevskii theory of
two-component BEC is closely related to the Skyrme theory,
which is really remarkable. 

The Hamiltonian (\ref{gpham1}) can be expressed as
\bea
&{\cal H} = \lambda \rho_0^4 ~{\hat {\cal H}}, \nn\\
&{\hat {\cal H}} = \dfrac {1}{2} (\hat \pro_i \hat \rho)^2 +
\dfrac {1}{2} \hat \rho^2 |\hat \pro_i \zeta|^2
+ \dfrac{1}{8} (\hat \rho^2-1)^2 \nn\\
&+ \dfrac{\delta \mu}{4\mu} \hat \rho^2 \zeta_2^*\zeta_2,
\label{gpham2}
\eea
where
\bea
&\hat \rho = \dfrac {\rho}{\rho_0},
~~~~~\hat \pro_i =\kappa \pro_i,
~~~~~\kappa = \dfrac {1}{\sqrt \lambda \rho_0}. \nn
\eea
Notice that ${\hat {\cal H}}$ is completely dimensionless, with
only one dimensionless coupling constant $\delta \mu/\mu$. This
tells that the physical unit of the Hamiltonian is $\lambda
\rho_0^4$, and the physical scale $\kappa$ of the coordinates is
$1/\sqrt \lambda \rho_0$. This is comparable to the correlation
length $\bar \xi$,
\bea
\bar \xi= \dfrac{1}{\sqrt {2\mu M}}
= \sqrt 2 ~\kappa.
\eea
For $^{87}{\rm Rb}$ we have
\bea
&M \simeq 8.1 \times 10^{10}~eV,
~~~~~\bar \lambda \simeq 1.68 \times 10^{-7}~(nm)^2, \nn\\
&\mu \simeq 3.3 \times 10^{-12}~eV, ~~~~~\delta \mu \simeq
0.1~\mu,
\label{data}
\eea
so that the density of $^{87}{\rm Rb}$ atom is given by
\bea
<\phi^{\dag} \phi> = \dfrac{\mu}{\bar \lambda}
\simeq 0.998 \times 10^{14}/cm^3.
\eea
From (\ref{data}) we have
\bea
&\lambda \simeq 1.14 \times 10^{11},
~~~~~\rho_0^2 \simeq 3.76 \times 10^{-11}~(eV)^2, \nn\\
&\delta \mu^2 \simeq 5.34 \times 10^{-2}~(eV)^2.
\eea
So the physical scale $\kappa$ for $^{87}{\rm Rb}$ becomes
about $1.84 \times 10^2~nm$.

\section{Vortex solutions in Gross-Pitaevskii Theory}

The two-component Gross-Pitaevskii theory is known to have
non-Abelian vortices \cite{met,bec5}.
To obtain the vortex solutions in two-component
Gross-Pitaevskii theory we first consider a straight vortex
with the ansatz
\bea
&\rho= \rho(\varrho),
~~~~~\zeta=\Bigg( \matrix{\cos \dfrac{f(\varrho)}{2} \exp (-in\varphi) \cr
\sin \dfrac{f(\varrho)}{2}} \Bigg).
\label{gpans}
\eea
With the ansatz (\ref{gpeq1}) is reduced to
\bea
&\ddot{\rho}+\dfrac{1}{\varrho}\dot{\rho}
-\bigg(\dfrac{1}{4}\dot{f}^2 + \dfrac{n^2}{\varrho^2}
-\big(\dfrac{n^2}{\varrho^2}
- \delta \mu^2 \big)\sin^2{\dfrac{f}{2}}\bigg)\rho \nn\\
&= \dfrac{\lambda}{2} (\rho^2-\rho_0^2) \rho,\nn \\
&\ddot{f}+\bigg(\dfrac{1}{\varrho}
+2\dfrac{\dot{\rho}}{\rho}\bigg)\dot{f}
+\bigg(\dfrac{n^2}{\varrho^2} - \delta \mu^2\bigg)\sin{f} \nn\\
&=0.
\label{gpeq3}
\eea
Now, we choose the following ansatz for a
helical vortex \cite{bec5}
\bea
&\rho= \rho(\varrho),
~~~~~\zeta = \Bigg( \matrix{\cos \dfrac{f(\varrho)}{2} \exp
(-in\varphi) \cr \sin \dfrac{f(\varrho)}{2} \exp (imkz)} \Bigg),
\label{gpans1}
\eea
and find that the equation (\ref{gpeq1}) becomes
\bea
&\ddot{\rho}+\dfrac{1}{\varrho}\dot{\rho}
-\bigg(\dfrac{1}{4}\dot{f}^2 + \dfrac{n^2}{\varrho^2} \nn\\
&-\big(\dfrac{n^2}{\varrho^2} -m^2 k^2 - \delta \mu^2
\big)\sin^2{\dfrac{f}{2}}\bigg)\rho
= \dfrac{\lambda}{2} (\rho^2-\rho_0^2) \rho,\nn \\
&\ddot{f}+\bigg(\dfrac{1}{\varrho}
+2\dfrac{\dot{\rho}}{\rho}\bigg)\dot{f}
+\bigg(\dfrac{n^2}{\varrho^2} - m^2 k^2 - \delta \mu^2\bigg)\sin{f} \nn\\
&=0.
\label{gpeq4}
\eea
Notice that mathematically this equation becomes identical to 
the equation of the straight vortex (\ref{gpeq3}), except that here
$\delta \mu^2$ is replaced by $\delta \mu^2+m^2 k^2$.

\begin{figure}
\includegraphics[scale=0.7]{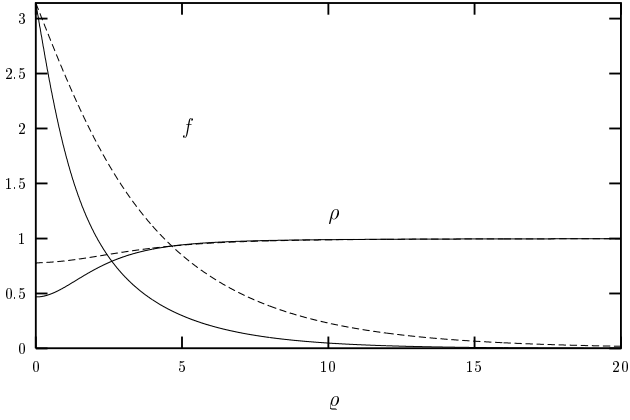}
\caption{The untwisted vortex in the Gross-Pitaevskii theory of
two-component BEC. Here we have put $n=1$, and $\varrho$ is in the
unit of $\kappa$. Dashed and solid lines correspond to $\delta
\mu/\mu= 0.1$ and $0.2$ respectively.}
\label{twobec-fig4}
\end{figure}

\begin{figure}
\includegraphics[scale=0.7]{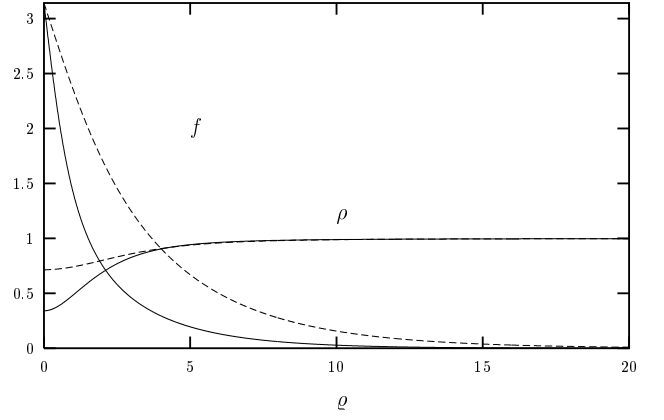}
\caption{The helical vortex in the Gross-Pitaevskii theory of
two-component BEC. Here we have put $m=n=1,~k=0.25 /\kappa$, and
$\varrho$ is in the unit of $\kappa$. Dashed and solid lines
correspond to $\delta \mu/\mu=0$, and $0.1$ respectively.}
\label{twobec-fig}
\end{figure}

Now, with the boundary condition
\bea
&\dot \rho(0)=0,~~~~~~\rho(\infty)=\rho_0,  \nn\\
&f(0)=\pi,~~~~~~f(\infty)=0,
\label{gpbc}
\eea
we can solve
(\ref{gpeq4}). With $m=0,~n=1$ we obtain the
straight (untwisted) vortex solution shown in Fig.
\ref{twobec-fig4}, but with $m=n=1$ we obtain the
twisted vortex solution shown in Fig. \ref{twobec-fig}.
Of course (\ref{gpeq1}) also admits the well-known Abelian vortices with
$\zeta_1=0$ or $\zeta_2=0$. But obviously they are different from
the non-Abelian vortices discussed here.

The untwisted non-Abelian vortex solution has been discussed
before \cite{gar,met}, but the twisted vortex
solution here is new \cite{bec5}.
Although they look very similar on the surface, they are
quite different. First, when $\delta \mu^2=0$, there is no untwisted
vortex solution because in this case the vortex size (the penetration
length of the vorticity) becomes infinite. However, the
helical vortex exists even when $\delta \mu^2=0$. This is because
the twisting reduces the size of vortex tube. More importantly,
they are physically different. The untwisted vortex is made of
a single vorticity flux, but the helical vortex is made of
two vorticity fluxes linked together \cite{bec5}.

In Skyrme theory the helical vortex is interpreted as a twisted
magnetic vortex, whose flux is quantized due to the topological
reason. The helical vortex in Gross-Pitaevskii theory is
also topological, which can be viewed as a quantized
vorticity flux \cite{bec5}. To see this
notice that, up to the overall factor two, the potential $C_\mu$ 
introduced in (\ref{hn}) is nothing but the velocity potential 
$V_\mu$ (more precisely the momentum potential) of the doublet
$\zeta$ \cite{bec1,bec5}
\bea
&V_\mu = -i\zeta^{\dagger} \pro_\mu \zeta
=\dfrac{1}{2} C_\mu \nn\\
&=-\dfrac{n}{2}(\cos{f}+1) \pro_\mu \varphi
-\dfrac{mk}{2}(\cos{f}-1) \pro_\mu z,
\label{gpvel}
\eea
which generates the vorticity
\bea
&\bar H_{\mu\nu}= \pro_\mu V_\nu - \pro_\nu V_\mu
=\dfrac{1}{2} H_{\mu\nu} \nn\\
&=\dfrac{\dot{f}}{2} \sin{f}\Big(n(\pro_\mu \varrho \pro_\nu
\varphi
-\pro_\nu \varrho \pro_\mu \varphi) \nn\\
&+mk(\pro_\mu \varrho \pro_\nu z - \pro_\nu \varrho \pro_\mu z)
\Big).
\label{gpvor}
\eea
This has two quantized vorticity fluxes,
$\Phi_{\hat z}$ along the $z$-axis
\bea
&\Phi_{\hat z}=\dfrac{}{}
\int \bar H_{{\hat \varrho}{\hat \varphi}} \varrho d \varrho d
\varphi = -2\pi n,
\label{gpfluxz}
\eea
and $\Phi_{\hat \varphi}$
around the $z$-axis (in one period section from $z=0$ to
$z=2\pi/k$)
\bea
&\Phi_{\hat \varphi}=\dfrac{}{} \int_0^{2\pi/k}
\bar H_{{\hat z}{\hat \varrho}} d \varrho dz = 2\pi m.
\label{gpfluxphi}
\eea
Clearly two fluxes are linked together.

\begin{figure}
\includegraphics[scale=0.5]{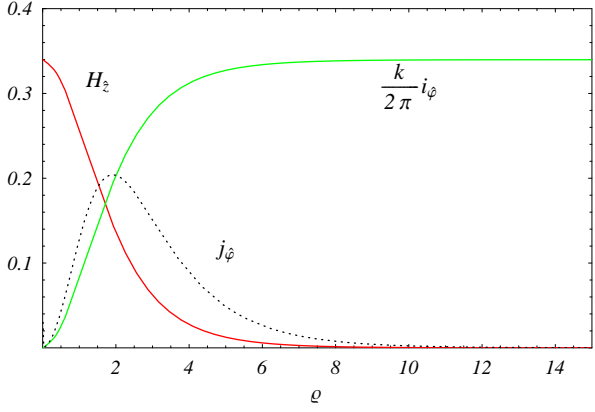}
\caption{The supercurrent $i_{\hat \varphi}$ (in one period
section in $z$-coordinate) and corresponding magnetic field
$H_{\hat z}$ circulating around the cylinder of radius $\varrho$
of the helical vortex in the Gross-Pitaevskii theory of
two-component BEC. Here $m=n=1,~k=0.25/\kappa,~\delta \mu^2=0$,
and $\varrho$ is in the unit of $\kappa$. The current density
$j_{\hat \varphi}$ is represented by the dotted line.}
\label{beciphi-fig}
\end{figure}

\begin{figure}
\includegraphics[scale=0.5]{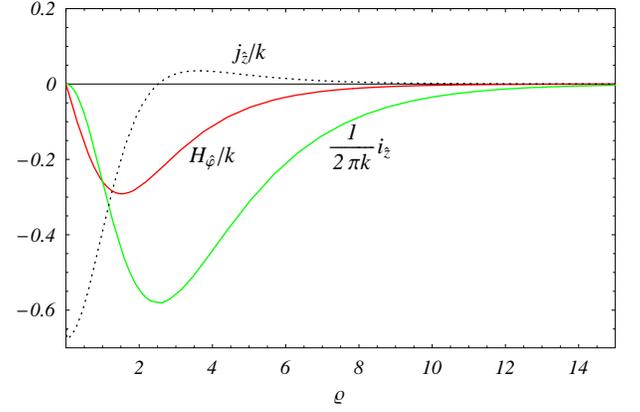}
\caption{The supercurrent $i_{\hat z}$ and corresponding magnetic
field $H_{\hat \varphi}$ flowing through the disk of radius
$\varrho$ of the helical vortex in the Gross-Pitaevskii theory of
two-component BEC. Here $m=n=1,~k=0.25 /\kappa,~\delta \mu^2=0$,
and $\varrho$ is in the unit of $\kappa$. The current density
$j_{\hat z}$ is represented by the dotted line.} \label{beciz-fig}
\end{figure}

Furthermore, just as in Skyrme theory, these fluxes can be viewed
to originate from the helical supercurrent which confines them
with a built-in Meissner effect
\bea
&j_\mu = \pro_\nu \bar H_{\mu\nu} \nn\\
&=-\sin f \Big[n \big(\ddot f + \dfrac{\cos f}{\sin f}
\dot f^2 - \dfrac{1}{\varrho} \dot f \big) \partial_{\mu}\varphi \nn\\
&+mk \big(\ddot f + \dfrac{\cos f}{\sin f} \dot f^2
+ \dfrac{1}{\varrho} \dot f \big) \partial_{\mu}z \Big], \nn\\
&\pro_\mu j_\mu = 0.
\label{gpsc}
\eea
This produces the supercurrents $i_{\hat\varphi}$ (per one period
section in $z$-coordinate from $z=0$ to $z=2\pi/k$) around the $z$-axis
\bea
&i_{\hat\varphi} = -\dfrac{2 \pi n}{k}\dfrac{\sin{f}}{\varrho}\dot
f \Bigg|_{\varrho=0}^{\varrho=\infty},
\eea
and $i_{\hat z}$ along the $z$-axis
\bea
&i_{\hat z} = -2 \pi mk \varrho \dot f \sin{f}
\Bigg|_{\varrho=0}^{\varrho=\infty}.
\eea
The vorticity fluxes and
the corresponding supercurrents are shown in Fig.
\ref{beciphi-fig} and Fig. \ref{beciz-fig}. This shows that
the helical vortex is made of two quantized vorticity fluxes,
the $\Phi_{\hat z}$ flux centered at the core and
the $\Phi_{\hat \varphi}$ flux surrounding it \cite{bec5}.
This is almost identical to what we have in Skyrme theory.
Indeed the remarkable similarity between Fig. \ref{beciphi-fig}
and Fig. \ref{beciz-fig} and  Fig. 4 and
Fig. 5 in Skyrme theory is unmistakable.
This confirms that the helical vortex of
two-component BEC is nothing but
two quantized vorticity fluxes linked together.
We emphasize that this interpretation holds even
when the $\delta \mu^2$ is not zero.

The quantization of the vorticity (\ref{gpfluxz}) and
(\ref{gpfluxphi}) is due to the non-Abelian topology
of the theory. To see this notice that the vorticity (\ref{gpvor})
is completely fixed by the non-linear sigma field $\hn$
defined by $\zeta$. Moreover,
for the straight vortex $\hn$
naturally defines a mapping $\pi_2(S^2)$ from the compactified
two-dimensional space $S^2$ to the target space $S^2$. 
This means that the vortex in two-component BEC has exactly
the same topological origin as the baby skyrmion in Skyrme theory.
The only difference is that the topological quantum number here
can also be expressed by the doublet $\zeta$
\bea
&Q_v = - \dfrac {i}{4\pi}
\int \epsilon_{ij} \partial_i \zeta^{\dagger}
\partial_j \zeta  d^2 x = n.
\label{gpvqn}
\eea
Exactly the same topology assures the quantization of
the twisted vorticity flux \cite{bec5}.
This clarifies the topological origin
of the non-Abelian vortices of
Gross-Pitaevskii theory in two-component BEC.

The helical vortex will become
unstable unless the periodicity condition is enforced by
hand. But just as in Skyrme theory we can make it
a stable knot by making it a twisted vortex ring smoothly connecting
two periodic ends. In this twisted vortex ring
the periodicity condition of the helical vortex
is automatically guaranteed, and the vortex ring
becomes a stable knot. In this knot
the $n$ flux $\Phi_{\hat z}$ winds
around $m$ flux $\Phi_{\hat \varphi}$ of the helical vortex.
Moreover the ansatz (\ref{gpans}) tells that $\Phi_{\hat z}$
is made of mainly the first component while $\Phi_{\hat \varphi}$
is made of mainly the second component of two-component BEC.
So physically the knot can be viewed as two vorticity fluxes
linked together, the one made of the first component and
the other made of the second component which surrounds it.

As importantly the very twist which causes the instability
of the helical vortex
now ensures the stability of the knot. This is so
because dynamically the momentum $mk$ along the $z$-axis
created by the twist now generates a net angular momentum
which provides the centrifugal repulsive force around the $z$-axis
preventing the knot to collapse.

Furthermore, this dynamical stability of the knot is now
backed up by the topological stability. Again this is because
the non-linear sigma field $\hn$, after forming a knot,
defines a mapping $\pi_3(S^2)$ from the compactified space $S^3$
to the target space $S^2$. So the knot acquires
a non-trivial topology $\pi_3(S^2)$ whose quantum number
is given by the Chern-Simon index of the
velocity potential,
\bea
&Q = - \dfrac {1}{4\pi^2} \int \epsilon_{ijk} \zeta^{\dagger}
\partial_i \zeta ( \partial_j \zeta^{\dagger}
\partial_k \zeta ) d^3 x \nn\\
&= \dfrac{1}{16\pi^2} \int \epsilon_{ijk} V_i \bar H_{jk} d^3x =mn.
\label{bkqn}
\eea
This is precisely the linking number of two
vorticity fluxes, which is formally identical to
the knot quantum number of Skyrme theory \cite{cho01,fadd1,bec5}.
This assures the topological stability of the knot, because two
fluxes linked together can not be disconnected by any smooth
deformation of the field configuration.

Similar knots in Gross-Pitaevskii theory of two-component BEC has
been discussed in the literature \cite{ruo,batt1}. Our analysis
here tells that the knot in Gross-Pitaevskii theory is 
a topological knot which can be viewed as a twisted
vorticity flux ring linked together.

As we have argued our knot should be stable, dynamically as well
as topologically. On the other hand the familiar scaling argument
indicates that the knot in Gross-Pitaevskii theory of
two-component BEC must be unstable. This has
created a confusion on the stability of the knot in the
literature \cite{ruo,met}. To clarify the confusion 
it is important to realize
that the scaling argument breaks down when the system is
constrained. In our case the Hamiltonian is constrained by the
particle number conservation which allows us to circumvents the
no-go theorem and have a stable knot \cite{met,bec5}.

\section{Gauge Theory of Two-component BEC}

The above analysis tells that the non-Abelian vortex of the
two-component Gross-Pitaevskii theory is nothing but a vorticity
flux. And creating the vorticity costs energy. This implies that
the Hamiltonian of two-component BEC must contain the contribution
of the vorticity. This questions the wisdom of the
Gross-Pitaevskii theory, because the Hamiltonian (\ref{gpham1})
has no such interaction. To make up this shortcoming
a gauge theory of two-component BEC which
can naturally accommodate the vorticity interaction has 
been proposed recently \cite{bec1}. In this section we discuss 
the gauge theory of two-component BEC in detail.

Let us consider the following Lagrangian of $U(1)$ gauge theory of
two-component BEC \cite{bec1}
\bea
&{\cal L} = i \dfrac {\hbar}{2}
\Big[\phi^\dag ( \tilde{D}_t \phi)
-( \tilde{D}_t \phi)^\dag \phi \Big] 
-\dfrac {\hbar^2}{2M} |\tilde{D}_i \phi|^2 \nn\\
& -\dfrac {\lambda}{2} \big(\phi^\dag \phi -\dfrac{\mu}{\lambda}
\big)^2 - \delta \mu \phi_2^\dag \phi_2 - \dfrac {1}{4}
\tilde{H}_{\mu \nu} ^2,
\label{beclag}
\eea
where
$\tilde{D}_\mu = \pro_\mu + i g \tilde{C}_\mu$, and
$\tilde{H}_{\mu\nu}$ is the field strength of the potential
$\tilde{C}_\mu$. Two remarks are in order here. First, from
now on we will assume
\bea
\delta \mu =0,
\label{dmu}
\eea
since the symmetry breaking interaction can always be treated as a
perturbation. With this the theory acquires a global $U(2)$
symmetry as well as a local $U(1)$ symmetry. Secondly,
since we are primarily interested in the self-interacting
(neutral) condensate, we treat the potential
$\tilde{C}_\mu$ as a composite field of the condensate
and identify $\tilde{C}_\mu$ with the velocity potential
$V_\mu$ of the doublet $\zeta$ \cite{bec1},
\bea
&\tilde{C}_\mu = -\dfrac{i}{g} \zeta^\dag \pro_\mu \zeta 
=\dfrac{1}{g} V_\mu.
\label{cm}
\eea
With this the last term in the Lagrangian
now represents the vorticity (\ref{gpvor}) of the
velocity potential that we discussed before
\bea
&\tilde{H}_{\mu \nu} = -\dfrac{i}{g} (\pro_\mu \zeta^\dag
\pro_\nu \zeta -\pro_\nu \zeta^\dag \pro_\mu \zeta)
=\dfrac{1}{g} \bar H_{\mu\nu}.
\label{hmn}
\eea
This shows that the gauge theory of two-component BEC naturally 
accommodates the vorticity interaction, and the coupling constant 
$g$ here describes the coupling strength of the vorticity 
interaction \cite{bec1}. This vorticity interaction
distinguishes the gauge theory from the Gross-Pitaevskii
theory.

At this point one might still wonder why one needs the vorticity in the
Lagrangian (\ref{beclag}), because in ordinary (one-component) BEC
one has no such interaction. The reason is that in ordinary BEC
the vorticity is identically zero, because there the velocity is
given by the gradient of the phase of the complex condensate. Only
a non-Abelian (multi-component) BEC can have a non-vanishing
vorticity. More importantly, it costs energy to create the
vorticity in non-Abelian superfluids \cite{ho}. So it is natural that the
two-component BEC (which is very similar to non-Abelian
superfluids) has the vorticity interaction.
Furthermore, here we can easily control the strength of the
vorticity interaction with the coupling constant $g$. Indeed, if
necessary, we could even turn off the vorticity interaction by
putting $g=\infty$. This justifies the presence of the vorticity
interaction in the Hamiltonian.

Another important difference between this theory and 
the Gross-Pitaevskii theory is the $U(1)$ gauge symmetry. 
Clearly the Lagrangian (\ref{beclag}) retains
the $U(1)$ gauge invariance, in spite of the
fact that the gauge field is replaced by the velocity field
(\ref{cm}). This has a deep impact. To see this 
notice that from the Lagrangian we have the following Hamiltonian
in the static limit (again normalizing $\phi$ to 
$\sqrt{2M/\hbar} \phi$)
\bea
&{\cal H} = \dfrac {1}{2} (\pro_i
\rho)^2 + \dfrac {1}{2} \rho^2 \Big(|\pro_i \zeta |^2 - |\zeta^\dag
\pro_i \zeta|^2 \Big)
+ \dfrac{\lambda}{2} (\rho^2 - \rho_0^2)^2 \nn\\
&- \dfrac {1}{4 g^2} (\pro_i \zeta^\dag \pro_j \zeta 
- \pro_j \zeta^\dag \pro_i \zeta)^2, \nn\\
&\rho_0^2= \dfrac{2\mu}{\lambda}.
\label{becham1}
\eea
Minimizing the Hamiltonian we have the following equation of
motion
\bea
& \pro^2 \rho - \Big(|\pro_i \zeta |^2 - |\zeta^\dag
\pro_i \zeta|^2 \Big)
\rho = \dfrac{\lambda}{2} (\rho^2 - \rho_0^2) \rho,\nn \\
&\Big\{(\pro^2 - \zeta^\dag \pro^2 \zeta) + 2 \Big(\dfrac {\pro_i
\rho}{\rho} + \dfrac {1}{g^2 \rho^2} \pro_j (\pro_i \zeta^\dag
\pro_j \zeta - \pro_j \zeta^\dag \pro_i \zeta) \nn\\
&- \zeta^\dag \pro_i\zeta \Big) (\pro_i 
- \zeta^\dag \pro_i \zeta) \Big\} \zeta = 0.
\label{beceq1}
\eea
But factorizing $\zeta$ by the $U(1)$ phase $\gamma$ and
$CP^1$ field $\xi$ with
\bea
\zeta= \exp(i\gamma) \xi,
\label{xi}
\eea
we have
\bea
&\zeta^\dag \vec \sigma \zeta = \xi^\dag \vec \sigma \xi = \hn, \nn\\ 
&|\pro_\mu \zeta |^2 - |\zeta^\dag \pro_\mu \zeta|^2 
=|\pro_\mu \xi |^2 - |\xi^\dag \pro_\mu \xi|^2 \nn\\
&=\dfrac {1}{4} (\pro_\mu \hn)^2, \nn\\
&-i(\pro_\mu \zeta^\dag \pro_\nu \zeta
- \pro_\nu \zeta^\dag \pro_\mu \zeta) 
=-i(\pro_\mu \xi^\dag \pro_\nu \xi
- \pro_\nu \xi^\dag \pro_\mu \xi) \nn\\ 
&=\dfrac {1}{2} \hn \cdot (\pro_\mu \hn \times \pro_\nu \hn)
= g \tilde H_{\mu\nu},
\label{xiid}
\eea
so that we can rewrite (\ref{beceq1}) in terms of $\xi$ 
\bea
& \pro^2 \rho - \Big(|\pro_i \xi |^2 - |\xi^\dag
\pro_i \xi|^2 \Big)
\rho = \dfrac{\lambda}{2} (\rho^2 - \rho_0^2) \rho,\nn \\
&\Big\{(\pro^2 - \xi^\dag \pro^2 \xi) + 2 \Big(\dfrac {\pro_i
\rho}{\rho} + \dfrac {1}{g^2 \rho^2} \pro_j (\pro_i \xi^\dag
\pro_j \xi
- \pro_j \xi^\dag \pro_i \xi) \nn\\
&- \xi^\dag \pro_i\xi \Big) (\pro_i - \xi^\dag \pro_i \xi) \Big\}
\xi = 0.
\label{beceq2}
\eea
Moreover we can express the Hamiltonian (\ref{becham1}) completely 
in terms of the non-linear sigma field $\hn$ (or equivalently 
the $CP^1$ field $\xi$) and $\rho$ as
\bea
&{\cal H} = \dfrac {1}{2} (\pro_i
\rho)^2 + \dfrac {1}{8} \rho^2 (\pro_i \hn)^2
+ \dfrac{\lambda}{2} (\rho^2 - \rho_0^2)^2 \nn\\
&+ \dfrac {1}{16 g^2} (\pro_i \hn \times \pro_j \hn)^2 \nn\\
&= \lambda \rho_0^4 \Big\{\dfrac {1}{2} (\hat \pro_i
\hat \rho)^2 + \dfrac {1}{8} \hat \rho^2 (\hat \pro_i \hn)^2
+ \dfrac{1}{2} (\hat \rho^2 - 1)^2 \nn\\
&+ \dfrac {\lambda}{16 g^2} (\hat \pro_i \hn \times \hat \pro_j \hn)^2.
\label{becham2}
\eea
This is because of the $U(1)$ gauge 
symmetry. The $U(1)$ gauge invariance of the Lagrangian (\ref{beclag})
absorbs the $U(1)$ phase $\gamma$ of $\zeta$, so that the theory
is completely described by $\xi$. In other words,
the Abelian gauge invariance of effectively reduces
the target space $S^3$ of $\zeta$ to the gauge orbit space $S^2 =
S^3/S^1$, which is identical to the target space of the $CP^1$
field $\xi$. And since mathematically $\xi$ is equivalent to the
non-linear sigma field $\hn$, one can
express (\ref{becham1}) completely in terms of $\hn$. 

This tells that the equation (\ref{beceq1}) can also be expressed 
in terms of $\hn$. Indeed with
(\ref{nid}), (\ref{cid}), and (\ref{xiid}) we can obtain
the following equation from (\ref{beceq1}) \cite{bec1}
\bea
&\pro^2 \rho - \dfrac{1}{4} (\pro_i
\n)^2 \rho = \dfrac{\lambda}{2} (\rho^2 -
\rho_0^2) \rho, \nn \\
&\n \times \pro^2 \n + 2 \dfrac{\pro_i \rho}{\rho} \n \times
\pro_i \n + \dfrac{2}{g \rho^2} \pro_i \tilde H_{ij} \pro_j \n \nn\\
&= 0.
\label{beceq3}
\eea
This, of course, is the equation of motion that one obtains 
by minimizing the Hamiltonian (\ref{becham2}). 
So we have two expressions, (\ref{beceq1}) and (\ref{beceq3}), 
which describe the equation of
gauge theory of two-component BEC.

The above analysis clearly shows that our theory of two-component
BEC is closely related to the Skyrme theory. In fact, in the
vacuum
\bea
\rho^2=\rho_0^2,
~~~~~\dfrac{1}{g^2 \rho_0^2}=\dfrac{\alpha}{\mu^2},
\label{vac}
\eea
the two equations (\ref{sfeq}) and (\ref{beceq3}) become identical.
Furthermore, this tells that the equation (\ref{gpeq2}) of
Gross-Pitaevskii theory is very similar to the above equation 
of gauge theory of two-component BEC.
Indeed, when $\delta \mu^2=0$, (\ref{gpeq2}) and
(\ref{beceq3}) become almost identical.
This tells that, in spite of different dynamics, the two theories
are very similar to each other.

\section{Topological Objects in Gauge Theory of Two-component BEC}

\begin{figure}[t]
\includegraphics[scale=0.7]{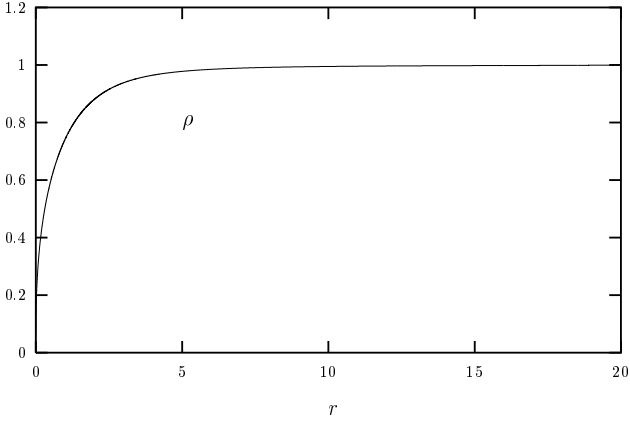}
\caption{The monopole solution in the gauge theory of
two-component BEC. Here we have put $\lambda=1$ and $r$ is in the
unit of $1/\rho_0$} \label{becmono}
\end{figure}

Now we show that, just like the Skyrme theory, the theory admits
monopole, vortex, and knot. We start from the monopole. Let
\bea
&\phi= \dfrac{1}{\sqrt 2} \rho \xi~~~~~(\gamma=0), \nn\\
&\rho = \rho(r),
~~~~~\xi = \Bigg( \matrix{\cos \dfrac{\theta}{2} \exp (-i\varphi) \cr
\sin \dfrac{\theta}{2} } \Bigg),
\eea
and find
\bea
&\hn =\xi^\dag
\vec \sigma \xi = \hat r.
\label{mono2}
\eea
where
$(r,\theta,\varphi)$ are the spherical coordinates. With this the
second equation of (\ref{beceq2}) is automatically satisfied, and
the first equation is reduced to
\bea
&\ddot \rho + \dfrac{2}{r}
\dot\rho - \dfrac{1}{2r^2} \rho = \dfrac{\lambda}{2} (\rho^2 -
\rho_0^2) \rho.
\eea
So with the boundary condition
\bea
\rho (0)=0,~~~~~\rho (\infty) = \rho_0,
\label{bmbc}
\eea
we have a spherically symmetric solution shown in 
Fig.~\ref{becmono}. Obviously this is a Wu-Yang type vorticity 
monopole dressed by the scalar field $\rho$ \cite{cho01,prd80}. 
In spite of the dressing, however, it has an infinite energy 
due to the singularity at the origin.

\begin{figure}[t]
\includegraphics[scale=0.5]{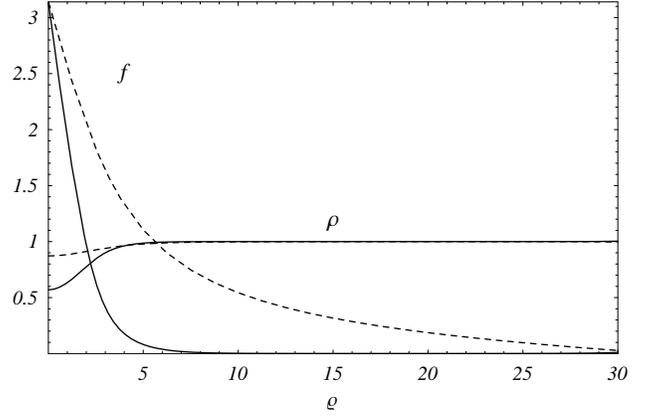}
\caption{The non-Abelian vortex (dashed line) with $m=0,n=1$ and
the helical vortex (solid line) with $m=n=1$ in the gauge theory
of two-component BEC. Here we have put $\lambda/g^2=1$,
$k=0.64/\kappa$, and $\varrho$ is in the unit of $\kappa$.}
\label{becheli}
\end{figure}

Next we construct the vortex solutions. To do
this we choose the ansatz in the cylindrical coordinates
\bea
&\rho= \rho(\varrho),
~~~~~\xi = \Bigg( \matrix{\cos \dfrac{f(\varrho)}{2} \exp (-in\varphi)
\cr \sin \dfrac{f(\varrho)}{2} \exp (imkz)}  \Bigg),
\eea
from
which we have
\bea
&\n=\Bigg(\matrix{\sin{f}\cos{(n\varphi+mkz)} \cr
\sin{f}\sin{(n\varphi+mkz)} \cr \cos{f}}\Bigg), \nn\\
&\tilde{C}_\mu = -\dfrac{n}{2g} (\cos{f} +1) \pro_\mu \varphi \nn\\
&-\dfrac{mk}{2g} (\cos{f} -1) \pro_\mu z.
\label{bhvans}
\eea
With this
the equation (\ref{beceq2}) is reduced to
\bea
&\ddot{\rho}+\dfrac{1}{\varrho}\dot\rho -
\dfrac{1}{4}\Big(\dot{f}^2+(m^2 k^2+\dfrac{n^2}{\varrho^2})
\sin^2{f}\Big)\rho \nn\\
&= \dfrac{\lambda}{2}(\rho^2-\rho_0^2)\rho, \nn\\
&\Big(1+(\dfrac{n^2}{\varrho^2}+m^2 k^2)
\dfrac{\sin^2{f}}{g^2 \rho^2}\Big) \ddot{f} \nn\\
&+ \Big( \dfrac{1}{\varrho}+ 2\dfrac{\dot{\rho}}{\rho}
+(\dfrac{n^2}{\varrho^2}+m^2 k^2)
\dfrac{\sin{f}\cos{f}}{g^2 \rho^2} \dot{f} \nn\\
&- \dfrac{1}{\varrho} (\dfrac{n^2}{\varrho^2}-m^2 k^2)
\dfrac{\sin^2{f}}{g^2 \rho^2} \Big) \dot{f} \nn\\
&- (\dfrac{n^2}{\varrho^2}+m^2 k^2) \sin{f}\cos{f}=0.
\label{bveq}
\eea
Notice that the first equation is similar to what we have in
Gross-Pitaevskii theory, but the second one is remarkably similar
to the helical vortex equation in Skyrme theory. Now with the
boundary condition
\bea
&\dot \rho(0)=0,~~~~~\rho(\infty)=\rho_0,  \nn\\
&f(0)=\pi,~~~~~f(\infty)=0,
\label{becbc}
\eea
we obtain the
non-Abelian vortex solution shown in Fig.~\ref{becheli}.

The solution is similar to the one we have in 
Gross-Pitaevskii theory. First, when $m=0$, the
solution describes the straight non-Abelian vortex. But when $m$
is not zero, it describes a helical vortex which is periodic in
$z$-coordinate \cite{bec1}. In this case, the vortex has a
non-vanishing velocity current (not only around the vortex but
also) along the $z$-axis. Secondly, the doublet $\xi$ starts from
the second component at the core, but the first component takes
over completely at the infinity. This is due to the boundary
condition $f(0)=\pi$ and $f(\infty)=0$, which assures that our
solution describes a genuine non-Abelian vortex. This confirms that the
vortex solution is almost identical to what we have in the
Gross-Pitaevskii theory shown in Fig. 5. All the qualitative
features are exactly the same. This implies that physically the
gauge theory of two-component BEC is very similar to the
Gross-Pitaevskii theory, in spite of the obvious dynamical
differences.

\begin{figure}[t]
\includegraphics[scale=0.5]{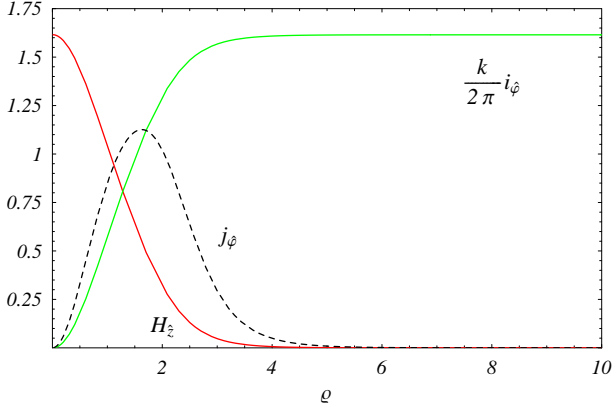}
\caption{The supercurrent $i_{\hat \varphi}$ (in one period
section in $z$-coordinate) and corresponding magnetic field
$H_{\hat z}$ circulating around the cylinder of radius $\varrho$
of the helical vortex in the gauge theory of two-component BEC.
Here $m=n=1$, $\lambda/g^2=1,~k=0.64/\kappa$, and $\varrho$ is in
the unit of $\kappa$. The current density $j_{\hat \varphi}$ is
represented by the dotted line.} \label{beciphi}
\end{figure}

Clearly our vortex has the same topological origin as the vortex
in Gross-Pitaevskii theory. This tells that, just as in
Gross-Pitaevskii theory, the non-Abelian helical vortex here is
nothing but the twisted vorticity flux of the $CP^1$ field $\xi$
confined along the $z$-axis by the velocity current, whose flux is
quantized due to the topological reason. The only difference here
is the profile of the vorticity, which is slightly different from
that of the Gross-Pitaevskii theory. Indeed the solution has the
following vorticity
\bea
&\tilde{H}_{\hat{z}}=\dfrac{1}{\varrho}\tilde{H}_{\varrho\varphi}
=\dfrac{n}{2g\varrho}\dot{f}\sin{f}, \nn\\
&\tilde{H}_{\hat{\varphi}}=-\tilde{H}_{\varrho
z}=-\dfrac{mk}{2g}\dot{f}\sin{f}, 
\eea 
which gives two quantized
vorticity fluxes, a flux along the $z$-axis 
\bea
&\phi_{\hat z} = \dfrac {}{}\int 
\tilde{H}_{\varrho\varphi} d\varrho d\varphi \nn\\
& = -\dfrac {2\pi i}{g} \int (\partial_{\varrho} \xi^{\dagger}
\partial_{\varphi} \xi - \partial_{\varphi} \xi^{\dagger}
\partial_{\varrho} \xi) d\varrho 
= -\dfrac{2\pi n}{g}, 
\label{nqn}
\eea
and a flux around the $z$-axis (in one period section from $0$ to
$2\pi/k$ in $z$-coordinate)
\begin{eqnarray}
&\phi_{\hat \varphi} = -\dfrac {}{}\int 
\tilde{H}_{\varrho z} d\varrho dz \nn\\
&= \dfrac {2\pi i}{g} \int (\partial_{\varrho} \xi^{\dagger}
\partial_z \xi - \partial_z \xi^{\dagger}
\partial_{\varrho} \xi) \dfrac{d\varrho}{k} 
= \dfrac{2\pi m}{g}.
\label{mqn}
\end{eqnarray}
This tells that the vorticity
fluxes are quantized in the unit of $2\pi/g$.

\begin{figure}[t]
\includegraphics[scale=0.5]{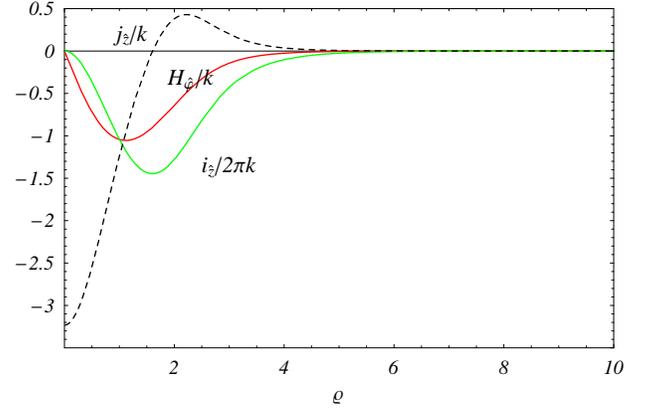}
\caption{The supercurrent $i_{\hat z}$ and corresponding magnetic
field $H_{\hat \varphi}$ flowing through the disk of radius
$\varrho$ of the helical vortex in the gauge theory of
two-component BEC. Here $m=n=1$, $\lambda/g^2=1,~k=0.64/\kappa$,
and $\varrho$ is in the unit of $\kappa$. The current density
$j_{\hat z}$ is represented by the dotted line.} \label{beciz}
\end{figure}

Just like the Gross-Pitaevskii theory the theory has a built-in
Meissner effect which confines the vorticity flux.
The current which confines the flux is given by
\bea
&j_\mu = \pro_{\nu} \tilde{H}_{\mu\nu} \nn\\
&=-\dfrac{\sin f}{2g} \Big[n \big(\ddot f + \dfrac{\cos f}{\sin f}
\dot f^2 - \dfrac{1}{\varrho} \dot f \big) \partial_{\mu}\varphi \nn\\
&+mk \big(\ddot f + \dfrac{\cos f}{\sin f} \dot f^2 +
\dfrac{1}{\varrho} \dot f \big) \partial_{\mu}z \Big], \nn\\
&\partial_{\mu} j_\mu = 0.
\eea
This produces the supercurrents $i_{\hat\varphi}$ (per one period
section in $z$-coordinate from $z=0$ to $z=2\pi/k$) around the
$z$-axis
\bea
&i_{\hat\varphi} = -\dfrac{\pi
n}{gk}\dfrac{\sin{f}}{\varrho}\dot f
\Bigg|_{\varrho=0}^{\varrho=\infty},
\eea
and $i_{\hat z}$ along
the $z$-axis
\bea
&i_{\hat z} = -\pi \dfrac{mk}{g} \varrho \dot f
\sin{f} \Bigg|_{\varrho=0}^{\varrho=\infty}.
\eea
The helical
vorticity fields and supercurrents are shown in Fig.~\ref{beciphi}
and Fig.~\ref{beciz}. The remarkable similarity between these and
those in Skyrme theory (Fig.~\ref{skyiphi} and Fig.~\ref{skyiz})
and Gross-Pitaevskii theory (Fig. \ref{beciphi-fig} and Fig.
\ref{beciz-fig}) is unmistakable.

With the ansatz (\ref{bhvans}) the energy (per one periodic section)
of the helical vortex is given by
\begin{eqnarray}
&E = \dfrac{4\pi^2}{k}\displaystyle\int^\infty_0 \bigg\{\dfrac{1}{2}
\dot{\rho}^2+ \dfrac{1}{8}\rho^2 \bigg( \big(1 \nn\\
&+\dfrac{1}{g^2 \rho^2} (\dfrac{n^2}{\varrho^2} + m^2k^2) \sin^2f
\big) \dot{f}^2 +  (\dfrac{n^2}{\varrho^2} \nn\\
&+m^2k^2)\sin^2{f} \bigg) +\dfrac{\lambda}{8}(\rho^2-\rho_0^2)^2
\bigg\}\varrho d\varrho \nn\\
&=4\pi^2 \dfrac{\rho_0^2}{k}\displaystyle\int^\infty_0 \bigg\{\dfrac{1}{2}
\big(\dfrac{d\hat \rho}{dx}\big)^2+ \dfrac{1}{8}\hat \rho^2 \bigg( \big(1 \nn\\
&+\dfrac{\lambda}{g^2 \hat \rho^2} (\dfrac{n^2}{x^2}
+ m^2\kappa^2 k^2) \sin^2f \big) \big(\dfrac{df}{dx}\big)^2 \nn\\
&+ (\dfrac{n^2}{x^2} +m^2\kappa^2 k^2)\sin^2{f} \bigg) \nn\\
&+\dfrac{1}{8}(\hat \rho^2-1)^2 \bigg\}x dx.
\end{eqnarray}
One could calculate the energy of the
helical vortex numerically. With $m=n=1$ and $k=0.64~\kappa$ we
find that the energy in one period section of the helical vortex
in $^{87}{\rm Rb}$ is given by
\bea
&E \simeq
51~\dfrac{\rho_0}{\sqrt \lambda} 
\simeq 4.5 \times 10^{-10}~eV \nn\\
&\simeq 0.7 MHz,
\label{chve2}
\eea
which will have an important meaning later.

\section{Vorticity Knot in Two-component BEC}

The existence of the helical vortex predicts the existence of 
a topological knot in the gauge theory of two-component BEC,
for exactly the same reason that the helical vortices 
in Skyrme theory and Gross-Pitaevskii theory assure the existence 
of knots in these theories. To demonstrate 
the existence of knot in the gauge theory
of two-component BEC we introduce the toroidal coordinates
$(\eta,\gamma,\varphi)$ defined by
\begin{eqnarray}
&x=\dfrac{a}{D}\sinh{\eta}\cos{\varphi},
~~~y=\dfrac{a}{D}\sinh{\eta}\sin{\varphi}, \nn\\
&z=\dfrac{a}{D}\sin{\gamma}, \nn\\
&D=\cosh{\eta}-\cos{\gamma}, \nn\\
&ds^2=\dfrac{a^2}{D^2} \Big(d\eta^2+d\gamma^2+\sinh^2\eta
d\varphi^2 \Big), \nn\\
&d^3x=\dfrac{a^3}{D^3} \sinh{\eta} d\eta d\gamma d\varphi,
\label{tc}
\end{eqnarray}
where $a$ is the radius of the knot defined by $\eta=\infty$.
Notice that in toroidal coordinates, $\eta=\gamma=0$  represents
spatial infinity of $R^3$, and $\eta=\infty$ describes
the torus center.

Now we choose the following ansatz,
\begin{eqnarray}
&\phi=\dfrac{\rho (\eta,\gamma)}{\sqrt 2} \Bigg(\matrix{\cos
\dfrac{f(\eta,\gamma )}{2} \exp (-in\omega (\eta,\gamma) ) \cr
\sin \dfrac{f(\eta ,\gamma)}{2} \exp (im\varphi)} \Bigg).
\label{bkans}
\end{eqnarray}
With this we have the velocity potential
\bea
&\tilde{C}_\mu = -\dfrac{m}{2g} (\cos f-1)\partial_\mu \varphi \nn\\
&- \dfrac{n}{2g} (\cos f+1) \partial_\mu \omega,
\label{kvp}
\eea
which generates the
vorticity
\begin{eqnarray}
&\tilde{H}_{\mu \nu}= \partial_\mu \tilde{C}_\nu 
-\partial_\nu \tilde{C}_\mu, \nn\\
&\tilde{H}_{\eta \gamma }=\dfrac{n}{2g} K \sin f,
~~~~~\tilde{H}_{\gamma \varphi }=\dfrac{m}{2g} \sin f\partial_\gamma f, \nn\\
&\tilde{H}_{\varphi \eta }=- \dfrac{m}{2g} \sin f\partial_\eta f, 
\label{kvf1}
\end{eqnarray}
where
\bea
K = \partial _\eta f\partial _\gamma \omega
-\partial _\gamma f \partial_\eta \omega.
\eea
Notice that, in the orthonormal frame
$(\hat \eta, \hat \gamma, \hat \varphi)$, we have
\begin{eqnarray}
&\tilde{C}_{\hat{\eta}}=- \dfrac{nD}{2ga}
(\cos f +1)\partial _\eta \omega, \nn\\
&\tilde{C}_{\hat{\gamma}}=- \dfrac {nD}{2ga}
(\cos f+1)\partial_\gamma \omega, \nn\\
&\tilde{C}_{\hat{\varphi}}=-\dfrac{mD}{2ga\sinh \eta}(\cos f-1),
\end{eqnarray}
and
\begin{eqnarray}
&\tilde{H}_{\hat{\eta}\hat{\gamma}}=\dfrac{nD^2}{2ga^2} K \sin f, \nn\\
&\tilde{H}_{\hat{\gamma}\hat{\varphi}}=\dfrac{mD^2}{2ga^2\sinh
\eta} \sin f\partial_\gamma f, \nn\\
&\tilde{H}_{\hat{\varphi}\hat{\eta}}=- \dfrac{mD^2}{2ga^2\sinh
\eta} \sin f\partial_\eta f. 
\label{kvf2}
\end{eqnarray}

\begin{figure}[t]
\includegraphics[scale=0.5]{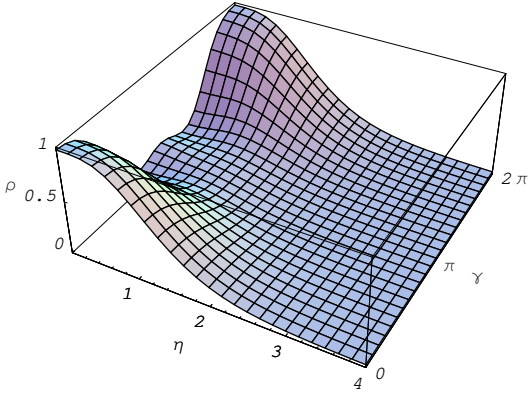}
\caption{(Color online). The $\rho$ profile of BEC knot with m = n
= 1. Here we have put $\lambda/g^2=1$, and the scale of the radius
$a$ is $\kappa$.}
\label{bkrho}
\end{figure}

From the ansatz (\ref{bkans}) we have the following equations of
motion
\begin{eqnarray}
&\Big[\partial_\eta^2+\partial_\gamma^2+(\dfrac{\cosh \eta}{\sinh \eta}
-\dfrac{\sinh \eta}D)\partial_\eta -\dfrac{\sin \gamma}D
\partial_\gamma \Big]\rho \nn\\
&-\dfrac14 \Big[(\partial _\eta f)^2
+(\partial_\gamma f)^2 \nn\\
&+ \Big(n^2 \big((\partial_\eta \omega)^2
+(\partial _\gamma \omega)^2 \big) 
+\dfrac{m^2}{\sinh ^2\eta}\Big) \sin^2 f \Big]\rho \nn\\
&=\dfrac{\lambda a^2}{2D^2}\Big(\rho^2-\rho_0^2\Big)\rho, \nn\\
&\Big[\partial_\eta^2 +\partial_\gamma^2
+\Big(\dfrac{\cosh \eta}{\sinh \eta}
-\dfrac{\sinh \eta}D\Big)\partial_\eta
-\dfrac{\sin \gamma}D\partial_\gamma \Big]f \nn\\
&-\Big(n^2 \big((\partial_\eta \omega)^2
+(\partial_\gamma \omega)^2 \big) +\dfrac{m^2}{\sinh ^2\eta}\Big)
\sin f\cos f \nn \\
&+\dfrac 2\rho \Big(\partial_\eta \rho \partial_\eta f
+\partial_\gamma \rho \partial_\gamma f\Big) \nn\\
&=-\dfrac 1{g^2\rho^2} \dfrac{D^2}{a^2}
\Big(A\cos f +B\sin f \Big)\sin f, \nn\\
&\Big[\partial_\eta^2 +\partial_\gamma^2
+(\dfrac{\cosh \eta}{\sinh \eta}-\dfrac{\sinh \eta}D)\partial _\eta
-\dfrac{\sin \gamma}D\partial_\gamma \Big]\omega \nn\\
&+2\Big(\partial_\eta f\partial_\eta \omega
+\partial_\gamma f\partial_\gamma \omega \Big)
\dfrac{\cos f}{\sin f} \nonumber \\
&+\dfrac 2\rho \Big(\partial_\eta \rho \partial_\eta \omega
+\partial_\gamma \rho \partial_\gamma \omega \Big) \nn\\
&=\dfrac1{g^2\rho^2} \dfrac{D^2}{a^2} C,
\label{bkeq1}
\end{eqnarray}
where
\begin{eqnarray}
&A=n^2 K^2+\dfrac{m^2}{\sinh ^2\eta}
\Big((\partial_\eta f)^2+(\partial _\gamma f)^2\Big), \nn\\
&B = n^2 \partial_\eta K\partial_\gamma \omega -n^2
\partial_\gamma K\partial_\eta \omega \nn\\
&+ n^2 K\Big[(\dfrac{\cosh \eta
}{\sinh \eta } +\dfrac{\sinh \eta }D)\partial_\gamma
\omega-\dfrac{\sin \gamma}D
\partial_\eta \omega \Big] \nn\\
&+\dfrac{m^2}{\sinh^2\eta}\Big[\partial_\eta^2 +\partial_\gamma^2 \nn\\
&-(\dfrac{\cosh \eta}{\sinh \eta} -\dfrac{\sinh \eta}D)
\partial_\eta +\dfrac{\sin \gamma}D
\partial_\gamma \Big]f , \nn\\
&C=\partial_\eta K\partial_\gamma f -\partial_\eta
f\partial_\gamma K \nn\\
&+K\Big[(\dfrac{\cosh \eta}{\sinh
\eta}+\dfrac{\sinh \eta}D)
\partial_\gamma -\dfrac{\sin \gamma}D\partial_\eta \Big]f. \nn
\end{eqnarray}

\begin{figure}[t]
\includegraphics[scale=0.5]{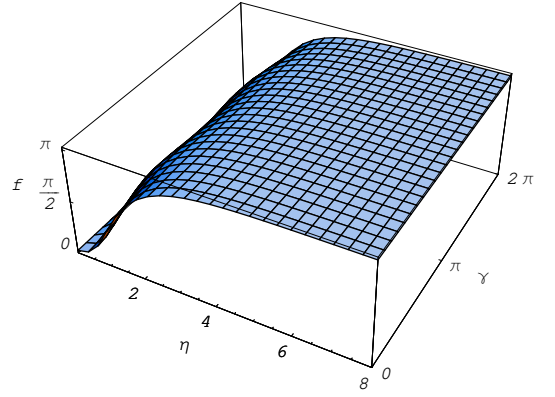}
\caption{(Color online). The $f$ profile of BEC knot
with $m=n=1$. Here we have put $\lambda/g^2=1$.}
\label{bkf}
\end{figure}

Since $\eta=\gamma=0$  represents spatial
infinity of $R^3$ and $\eta=\infty$ describes the torus center, we
can impose the following boundary condition
\begin{eqnarray}
&\rho(0,0)=\rho_0,
~~~~~\dot \rho(\infty,\gamma)=0, \nn\\
&f(0,\gamma)=0,
~~~~~f(\infty,\gamma)=\pi, \nn\\
&\omega(\eta,0)=0,
~~~~~\omega(\eta,2 \pi)=2 \pi,
\label{bkbc}
\end{eqnarray}
to obtain the desired knot.
From the ansatz (\ref{bkans}) and the boundary condition
(\ref{bkbc}) we can calculate the knot quantum number
\bea
&Q=\dfrac{mn}{8\pi ^2}\int K \sin f
d\eta d\gamma d\varphi \nn\\
&=  \dfrac{mn}{4\pi} \int \sin f df d\omega = mn,
\label{kqn10}
\eea
where the last equality comes from the
boundary condition. This tells that our ansatz describes the
correct knot topology.

\begin{figure}[t]
\includegraphics[scale=0.5]{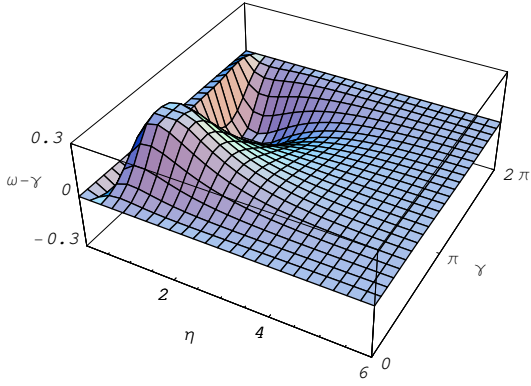}
\caption{(Color online). The $\omega$ profile of BEC knot
with $m=n=1$. Notice that here we have plotted $\omega-\gamma$.
Here again we have put $\lambda/g^2=1$.}
\label{bkdo}
\end{figure}

Of course, an exact solution of (\ref{bkeq1}) with the boundary
conditions (\ref{bkbc}) is extremely difficult to find
\cite{fadd1,batt1}. But here we can obtain the knot
profile of $\rho,~f$, and $\omega$ which minimizes 
the energy numerically. We find that, for $m=n=1$, 
the radius of knot which minimizes the
energy is given by
\bea
a \simeq 1.6 \kappa.
\label{bkrad}
\eea
From this we obtain the following solution of the lightest axially
symmetric knot in the gauge theory of two-component 
BEC (with $m=n=1$) shown in Fig.~\ref{bkrho},
Fig.~\ref{bkf}, and Fig.~\ref{bkdo}. With this we can obtain a
three-dimensional energy profile of the lightest knot 
(Unfortunately we can not show the profile here because 
the volume of the eps-file is too big).

We can calculate the vorticity flux of the knot. Since the
flux is helical, we have two fluxes, the flux $\Phi_{\hat \gamma}$
passing through the knot disk of radius $a$ in the $xy$-plane
and the flux $\Phi_{\hat \varphi}$ which surrounds it.
From (\ref{kvf2}) we have
\begin{eqnarray}
&\Phi_{\hat{\gamma}} = \dfrac{}{} \int_{\gamma=\pi}
\tilde{H}_{\hat{\gamma}}
\dfrac{a^2\sinh \eta}{D^2}d\eta d\varphi  \nn\\
&=- \dfrac{m}{2g}\int_{\gamma=\pi} \sin f\partial_\eta f d\eta d\varphi 
=-\dfrac{2\pi m}g,
\end{eqnarray}
and
\begin{eqnarray}
&\Phi_{\hat{\varphi}} = \dfrac{}{} \int \tilde{H}_{\hat{\varphi}}
\dfrac{a^2}{D^2}d\eta d\gamma  \nn\\
&=\dfrac{n}{2g} \int K \sin fd\eta d\gamma =\dfrac{2\pi n}g.
\label{kflux}
\end{eqnarray}
This confirms that the flux is quantized in the unit of $2\pi/g$.
As importantly this tells that the two fluxes are linked,
whose linking number is fixed by the knot quantum number.

Just as in Gross-Pitaevskii theory the vorticity flux here is
generated by the helical vorticity current which is conserved
\begin{eqnarray}
&j_\mu =\dfrac{nD^2}{2ga^2}\sin f \Big(\partial_\gamma 
+\dfrac{\sin \gamma }D\Big)K \partial_\mu \eta \nn\\
&-\dfrac{nD^2}{2ga^2}\sin f\Big(\partial_\eta
+\frac{\cosh \eta} {\sinh \eta}
+\frac{\sinh \eta}D\Big) K \partial_\mu \gamma \nn\\
&-\dfrac{mD^2}{2ga^2}\Big[\Big(\partial_\eta
-\frac{\cosh \eta }{\sinh \eta}
+\frac{\sinh \eta }D\Big)\sin f\partial_\eta f \nn\\
&+\Big(\partial_\gamma +\dfrac{\sin \gamma}D\Big)\sin f
\partial_\gamma f\Big] \partial_\mu \varphi,  \nn\\
&\pro_\mu j_\mu =0.
\label{kcd}
\end{eqnarray}
Clearly this supercurrent generates a Meissner
effect which confines the vorticity flux.

From (\ref{becham1}) and (\ref{bkans}) we have the following
Hamiltonian for the knot
\begin{eqnarray}
&{\cal H}= \dfrac{D^2}{2a^2}\Big\{(\partial_\eta \rho)^2
+(\partial_\gamma \rho)^2 \nn\\
&+\dfrac{\rho^2}4\Big[(\partial_\eta f)^2+(\partial_\gamma f)^2 
+\Big(n^2 \big((\partial _\eta \omega )^2
+(\partial_\gamma \omega)^2 \big) \nn\\
&+\dfrac{m^2}{\sinh^2 \eta}\Big)\Big] \sin^2 f \Big\}
+\dfrac{\lambda}{8} (\rho^2-\rho_0^2)^2 \nn\\
&+\dfrac{D^4}{8g^2a^4} A \sin^2 f.
\label{bkh}
\end{eqnarray}
With this the energy of the knot is given by
\begin{eqnarray}
&E=\dfrac{}{} \int {\cal H} \dfrac{a^3}{D^3}
\sinh \eta d\eta d\gamma d\varphi \nn\\
&=\dfrac{\rho_0}{\sqrt \lambda} \int {\hat {\cal H}} 
\dfrac{a^3}{\kappa^3 D^3} \sinh \eta d\eta d\gamma d\varphi,
\label{bke}
\end{eqnarray}
where
\bea
&{\hat {\cal H}}= \dfrac{\kappa^2 D^2}{2a^2}
\Big\{(\partial_\eta \hat \rho)^2
+(\partial_\gamma \hat \rho)^2 \nn\\
&+\dfrac{\hat \rho^2}4 \Big[(\partial_\eta f)^2+(\partial_\gamma f)^2 
+\Big(n^2 \big((\partial _\eta \omega )^2
+(\partial_\gamma \omega)^2 \big) \nn\\
&+\dfrac{m^2}{\sinh^2 \eta}\Big)\Big] \sin^2 f \Big\} 
+\dfrac{1}{8} (\hat \rho^2-1)^2 \nn\\
&+\dfrac{\lambda}{8g^2} \dfrac{\kappa^4 D^4}{a^4} A \sin^2 f.
\label{bkh1}
\eea
Minimizing the energy we reproduce the knot equation (\ref{bkeq1}).

%\begin{figure}[t]
%\includegraphics[scale=0.5]{beck3d.eps}
%\caption{(Color online). The energy profile of BEC knot with $m=n=1$
%for the energetically stable vorticity knot. Here again we have put
%$\sqrt \lambda/g^2=1$.}
%\label{bk3d}
%\end{figure}

\begin{figure}
\includegraphics[scale=0.6]{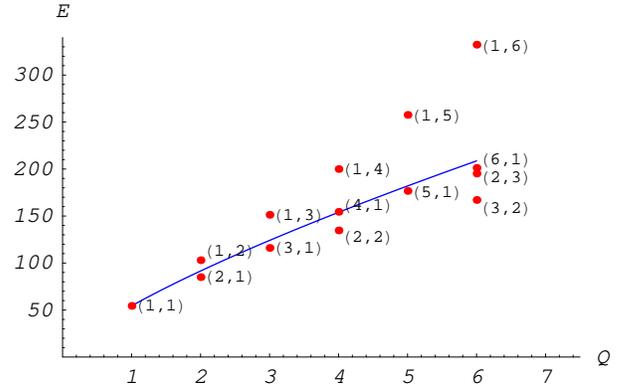}
\caption{(Color online). The $Q$-dependence of axially symmetric
knot. The solid line corresponds to the function
$E_0 Q^{3/4}$ with $E_0=E(1,1)$, and the red dots represent the
energy $E(m,n)$ with the different $Q=mn$.}
\label{bkqe}
\end{figure}

From this we can estimate the energy of the axially symmetric
knots. For the lightest knot (with $m=n=1$) we find the following
energy
\bea
&E \simeq 54~\dfrac{\rho_0}{\sqrt \lambda} \simeq
4.8\times 10^{-10} eV \nn\\
&\simeq 0.75 MHz.
\label{bke10}
\eea
One should compare this
energy with the energy of the helical vortex (\ref{chve2}). Notice
that the lightest knot has the radius $r \simeq 1.6 \kappa$. In
our picture this knot can be constructed bending a helical vortex
with $k \simeq 0.64/\kappa$. So we expect that the energy of the
lightest knot should be comparable to the energy of the helical
vortex with $k \simeq 0.64/\kappa$. And we have already estimated
the energy of the helical vortex with $k \simeq 0.64/\kappa$ in
(\ref{chve2}). The fact that two energies are of the same order 
assures that the knot can indeed be viewed as a twisted 
vorticity flux ring.

As we have remarked the $Q$-dependence of the energy of 
Faddeev-Niemi knot is proportional to $Q^{3/4}$ \cite{ussr,batt2}.
An interesting question is whether we can have a similar
$Q$-dependence of energy for the knots in BEC.
With our ansatz we have estimated the energy
of knot numerically for different $m$ and $n$ up to $Q=6$.
The result is summarized in Fig.~\ref{bkqe},
which clearly tells that the energy
depends crucially on $m$ and $n$.
Our result suggests that, for the minimum energy knots,
we have a similar (sub-linear) $Q$-dependence of energy 
for the knots in two-component BEC.
It would be very interesting
to establish such $Q$-dependence of energy mathematically.

\section{Discussion}

In this paper we have discussed two competing theories of
two-component BEC, the popular Gross-Pitaevskii theory and
the $U(1)$ gauge theory which has the vorticity interaction.
Although dynamically different two theories have remarkably
similar topological objects, the helical vortex and the knots,
which have a non-trivial non-Abelian topology.

We have shown that the $U(1)\times U(1)$ symmetry of two-component BEC
can be viewed as a broken $U(2)$ symmetry. This allows us to interpret
the vortex and knot in two-component BEC as non-Abelian topological
objects. Furthermore, we have shown that these topological objects are
the vorticity vortex and vorticity knot.

A major difference between the Gross-Pitaevskii theory and the
gauge theory is the vorticity interaction. In spite of the fact
that the vorticity plays an important role in two-component BEC,
the Gross-Pitaevskii theory has no vorticity interaction. In
comparison, the gauge theory of two-component BEC naturally
accommodates the vorticity interaction in the Lagrangian. This
makes the theory very similar to Skyrme theory. More
significantly, the explicit $U(1)$ gauge symmetry makes it very
similar to the theory of two-gap superconductor. The only
difference is that the two-component BEC is a neutral system which
is not charged, so that the gauge interaction has to be 
an induced interaction. On the other hand the two-gap 
superconductor is made of charged condensates, so that 
it has a real (independent)
electromagnetic interaction \cite{cm2}.

As importantly the gauge theory of two-component BEC, 
with the vorticity interaction, could play an important
role in describing multi-component superfluids \cite{bec1,ho}. In
fact we believe that the theory could well describe both
non-Abelian BEC and non-Abelian
superfluids. 

In this paper we have constructed a numerical solution of knot
in the gauge theory of two-component BEC. Our result confirms that
it can be identified as a vortex ring made of a helical vorticity
vortex. Moreover our result tells
that the knot can be viewed as two quantized vorticity fluxes
linked together, whose linking number becomes the knot quantum number.
This makes the knot very similar to Faddeev-Niemi knot in Skyrme theory.

We close with the following remarks: \\
1. Recently a number of authors have also established the
existence of knot identified as the ``skyrmions" in
Gross-Pitaevskii theory of two-component BEC \cite{ruo,batt1}, 
which we believe is identical to our knot in
Gross-Pitaevskii theory. 
In this paper we have clarified the physical meaning of the knot.
The knot in Gross-Pitaevskii theory is also of topological origin.
Moreover, it can be identified as a vorticity knot,
a twisted vorticity flux ring, 
in spite of the fact that the Gross-Pitaevskii Lagrangian
has neither the velocity $\tilde{C}_\mu$ nor the
vorticity $\tilde{H}_{\mu\nu}$ which can be related to the
knot. \\
2. Our analysis tells that at the center of the topological vortex
and knot in two-component BEC lies the baby skyrmion and the
Faddeev-Niemi knot. In fact they are the prototype of the
non-Abelian topological objects that one can repeatedly encounter
in physics \cite{cho01,bec1,sky3,plb05}. This suggests that the Skyrme
theory could also play an important role in condensed matter
physics. Ever since Skyrme proposed his theory, the Skyrme theory
has always been associated to nuclear and/or high energy physics.
This has lead people to believe that the topological objects in
Skyrme theory can only be realized at high energy, at the $GeV$
scale. But our analysis opens up a new possibility for us to
construct them in a completely different environment, at the $eV$
scale, in two-component BEC \cite{bec1,sky3}. 
This is really remarkable. \\
3. From our analysis there should be no doubt that the non-Abelian
vortices and knots must exist in two-component BEC. If so,
the challenge now is to verify the existence of these topological
objects experimentally. Constructing the knots might not be a
simple task at present moment. But the construction of the
non-Abelian vortices could be rather straightforward, and might
have already been done \cite{exp2,exp3}. Identifying them,
however, may be a tricky business because the two-component BEC
can admit both the Abelian and non-Abelian vortices. To identify
them, one has to keep the following in mind. First, the
non-Abelian vortices must have a non-trivial profile of
$f(\varrho)$. This is a crucial point which distinguishes them
from the Abelian vortices. Secondly, the energy of the non-Abelian
vortices must be bigger than that of the Abelian counterparts,
again because they have extra energy coming from the non-trivial
profile of $f$. With this in mind, one should be able to construct
the non-Abelian vortices in the new condensates without much
difficulty. We strongly urge the experimentalists to meet the
challenge.

{\bf ACKNOWLEDGEMENT}

One of us (YMC) thank G. Sterman for the kind hospitality during
his visit at C.N. Yang Institute for Theoretical Physics. The work
is supported in part by the ABRL Program of Korea Science and
Engineering Foundation (Grant R02-2003-000-10043-0), and by the
BK21 Project of the Ministry of Education.

\end{document}